\documentclass[10pt,aps,prd,onecolumn,showpacs,amsmath,amssymb,nofootinbib,eqsecnum,preprintnumbers,longbibliography,superscriptaddress]{revtex4-2}

\usepackage[english]{babel}				
\usepackage{mathtools}					
\usepackage{xcolor}			            

\usepackage[colorlinks=true,
            citecolor=red,
            linkcolor=blue,
            urlcolor=violet,
            filecolor=cyan,
            backref=false]{hyperref}

\newcommand\bs[1]{\boldsymbol{#1}}

\newcommand\dd{\mathrm{d}}
\newcommand\pp{\partial}

\newcommand\feq{\mathrel{\phantom{=}}}

\renewcommand{\Re}{\operatorname{Re}}
\renewcommand{\Im}{\operatorname{Im}}
\newcommand{\Ei}{\operatorname{Ei}}

\newcommand{\arccosh}{\operatorname{arccosh}}
\newcommand{\arctanh}{\operatorname{arctanh}}

\newcommand{\erf}{\operatorname{erf}}

\renewcommand{\ln}{\operatorname{log}}

\DeclareMathOperator*{\sumint}{%
\mathchoice%
  {\ooalign{$\displaystyle\sum$\cr\hidewidth$\displaystyle\int$\hidewidth\cr}}
  {\ooalign{\raisebox{.14\height}{\scalebox{.7}{$\textstyle\sum$}}\cr\hidewidth$\textstyle\int$\hidewidth\cr}}
  {\ooalign{\raisebox{.2\height}{\scalebox{.6}{$\scriptstyle\sum$}}\cr$\scriptstyle\int$\cr}}
  {\ooalign{\raisebox{.2\height}{\scalebox{.6}{$\scriptstyle\sum$}}\cr$\scriptstyle\int$\cr}}
}

\begin{document}


\title{Exact solutions of non-local gravity in a class of almost universal spacetimes}

\author{Ivan Kol\'a\v{r}}
\email{i.kolar@rug.nl}
\affiliation{Van Swinderen Institute, University of Groningen, 9747 AG, Groningen, Netherlands}

\author{Tom\'a\v{s} M\'alek}
\email{malek@math.cas.cz}
\affiliation{Institute of Mathematics of the Czech Academy of Sciences, \v{Z}itn\'a 25, 115 67 Prague 1, Czech Republic}

\author{Anupam Mazumdar}
\email{anupam.mazumdar@rug.nl}
\affiliation{Van Swinderen Institute, University of Groningen, 9747 AG, Groningen, Netherlands}

\date{\today}

\begin{abstract}
We study exact solutions of the infinite derivative gravity with null radiation which belong to the class of almost universal Weyl type III/N Kundt spacetimes. This class is defined by the property that all rank-2 tensors ${B_{ab}}$ constructed from the Riemann tensor and its covariant derivatives have traceless part of type N of the form $\mathcal{B}(\square)S_{ab}$ and the trace part constantly proportional to the metric. Here, $\mathcal{B}(\square)$ is an analytic operator and $S_{ab}$ is the traceless Ricci tensor. We show that the convoluted field equations reduce to a single non-local but linear equation, which contains only the Laplace operator $\triangle$ on 2-dimensional spaces of constant curvature. Such a non-local linear equation is always exactly solvable by eigenfunction expansion or using the heat kernel method for the non-local form-factor $\exp(-\ell^2\triangle)$ (with $\ell$ being the length scale of non-locality) as we demonstrate on several examples. We find the non-local analogues of the Aichelburg--Sexl and the Hotta--Tanaka solutions, which describe gravitational waves generated by null sources propagating in Minkowski, de Sitter, and anti-de Sitter spacetimes. They reduce to the solutions of the local theory far from the sources or in the local limit, ${\ell\to0}$. In the limit ${\ell\to\infty}$, they become conformally flat. We also discuss possible hints suggesting that the non-local solutions are regular at the locations of the sources in contrast to the local solutions; all curvature components in the natural null frame are finite and specifically the Weyl components vanish.
\end{abstract}

\maketitle


\section{Introduction}

Einstein's general relativity (GR) is a very successful theory at the scales of our solar system where it has surpassed many experimental tests. However, GR becomes problematic at short distances: i) its classical solutions suffer from spacetime singularities; ii) at the quantum level, it fails to be perturbatively renormalizable. An interesting resolution of these issues can be overcome with the \textit{non-locality}, which is present in many theories of quantum gravity. For example, in the string theory, the strings and branes interact over a certain region even at the classical level \cite{Polchinski:1998rr}. Similarly, the loop quantum gravity is inherently non-local since the quantization of space gives rise to a minimal area \cite{Ashtekar:2012np}. In fact, it proves to be almost impossible to define local variables in any theory of quantum gravity \cite{Kiefer:2014}. Thus, it is not surprising that all effective descriptions of string field theory \cite{Witten:1985cc,Freund:1987kt} and $p$-adic string theory \cite{Brekke:1988dg,Frampton:1988kr} feature non-local form-factors with an \textit{infinite number of derivatives}. The non-local short-distance completion of GR should occur at least at the Planck scale, but it is not ruled out experimentally anywhere below $50\,\mu\mathrm{m}$ scales \cite{PhysRevLett.124.101101}.

In the last two decades, a considerable interest has been attracted to a non-local gravity theory often referred to as the \textit{infinite derivative gravity} (IDG) \cite{Tomboulis:1997gg,Modesto:2011kw,Biswas:2011ar}. In a spacetime ${(M,\bs{g})}$ consisting of a 4-dimensional manifold $M$ equipped with the Lorentzian metric $\bs{g}$, it can be described by the action\footnote{A more common form of the action, ${R\tilde{\mathcal{F}}_1(\square)R+R_{ab}\tilde{\mathcal{F}}_2(\square)R^{ab}+C_{abcd}\tilde{\mathcal{F}}_3(\square)C^{abcd}}$, is related to \eqref{eq:action} via ${\tilde{\mathcal{F}}_1=\mathcal{F}_1-\mathcal{F}_2/4}$, ${\tilde{\mathcal{F}}_2=\mathcal{F}_2}$, and ${\tilde{\mathcal{F}}_3=\mathcal{F}_3}$.}
\begin{equation}\label{eq:action}
    S=\frac{1}{2}\int_M \! \mathfrak{g}^{\frac12} \, \Big[R-2\Lambda+R\mathcal{F}_1(\square)R+S_{ab}\mathcal{F}_2(\square)S^{ab}+C_{abcd}\mathcal{F}_3(\square)C^{abcd}\Big]+S_{\textrm{m}}\;,
\end{equation}
where $R$ denotes the Ricci scalar, $\bs{S}$ is the traceless Ricci tensor, $\bs{C}$ is the Weyl tensor, ${\mathfrak{g}^{\frac12}=\sqrt{-g}}\,dx^4$ stands for the volume element, $\square$ denotes the wave operator, and $\mathcal{F}_i(\square)$ are \textit{analytic} operators with constant coefficients $f_{i,n}$, 
\begin{equation}
    \mathcal{F}_i(\square)=\sum_{n=0}^{\infty}f_{i,n}\square^n\;.
\end{equation}
The infinite number of derivatives arise when the form-factors $\mathcal{F}_i(\square)$ are \textit{non-polynomial}. The theory \eqref{eq:action} contains a new parameter $\ell$ describing the \textit{length scale of non-locality}, which is hidden in the coefficients ${f_{i,n}=\ell^{2n+2}\hat{f}_{i,n}}$, where $\hat{f}_{i,n}$ are some dimensionless constants. In the \textit{local limit}, ${\ell\to 0}$, the action of IDG \eqref{eq:action} reduces to the standard Einstein--Hilbert action of GR with a cosmological constant $\Lambda$. To avoid ghost-like instabilities with respect to specific backgrounds \cite{Biswas:2010zk,Biswas:2013kla,Biswas:2016egy,SravanKumar:2019eqt}, the analytic form-factors are usually chosen as functions with no roots in the complex plane. The infinite number of derivatives obviously raises questions around non-perturbative degrees of freedom and an appropriate Hamiltonian formulation. There have been first attempts in understanding the initial value problem using the diffusion equation methods \cite{Calcagni:2018lyd,Calcagni:2018gke} and constructing Hamiltonians of scalar-field toy models \cite{Kolar:2020ezu} by shifting the non-locality into a new auxiliary dimension.

Due to the complicated nature of the theory, most of the research has focused on the \textit{weak field regime} of IDG. At the linearized level, it was shown that IDG can avoid various spacetime singularities: i) diverging Newton's potential is mollified by the error function \cite{Biswas:2011ar,Edholm:2016hbt,Buoninfante:2018rlq,Boos:2018bxf}; ii) there exists a mass gap for a mini-black-hole production in a collision of null sources \cite{Frolov:2015usa,Frolov:2015bia,Frolov:2015bta}; iii) solutions corresponding to topological defects such as the conical deficits \cite{Boos:2020kgj} and the Misner strings of NUT charges \cite{Kolar:2020bpo} are regularized at the axis; iv) fields of time-dependent \cite{Frolov:2016xhq} and uniformly accelerated sources \cite{Kolar:2021oba} are finite at the location of the source.

The \textit{exact} solutions of the full IDG are much scarcer in the literature due to the immense complexity of the field equations. Calculating the variation of the action \eqref{eq:action}, one can find \cite{Biswas:2013cha}
\begin{equation}\label{eq:fieldeq}
\begin{aligned}
E_{ab} \equiv{} & {S_{ab}}-\tfrac14 R{g_{ab}}+\Lambda{g_{ab}}+2{S_{ab}}\mathcal{F}_1(\square) R -2 \big({\nabla_a\nabla_b} -{g_{ab}}\square\big) \mathcal{F}_1(\square) R +\big(\square+\tfrac12 R\big)\mathcal{F}_2(\square) {S_{ab}}
\\
&-2{g_{d(a}}\big({\nabla^c}{\nabla^{d}}-{S^{cd}}\big) \mathcal{F}_2(\square) {S_{b)c}} +{g_{ab}} \big({\nabla^c} {\nabla^d}-\tfrac12{S^{cd}}\big) \mathcal{F}_2(\square) {S_{cd}}-4\big({\nabla^c}{\nabla^d}+\tfrac12{S^{cd}}\big)\mathcal{F}_3(\square){C_{d(ab)c}}
\\
&-{\Omega}_1{{}_{ab}}+\tfrac12 {g_{ab}}\big({\Omega}_1{{}^c{}_c}+\Theta_1\big)-{\Omega}_2{{}_{ab}}+\tfrac12 {g_{ab}}\big({\Omega}_2{{}^c{}_c}+\Theta_2\big)-{\Omega}_3{_{ab}}+\tfrac12 {g_{ab}}\big({\Omega}_3{{}^c{}_c}+\Theta_3\big)-2{\Upsilon}_2{{}_{ab}}-4{\Upsilon}_3{{}_{ab}}={T_{ab}}\;,
\end{aligned}
\end{equation}
where the tensors $\bs{\Omega}_i$, $\bs{\Upsilon}_i$, and scalars ${\Theta_i}$ are given by the double-infinite series
\begin{equation}\label{eq:LK}
\begin{gathered}
\begin{aligned}
    {\Omega}_1{{}_{ab}} &=\sum_{n=1}^\infty f_{1,n}\sum_{l=0}^{n-1}{\nabla_a} \square^l R\,{\nabla_b} \square^{n-l-1}R\;, & \Theta_1 &=\sum_{n=1}^\infty f_{1,n}\sum_{l=0}^{n-1}\square^l R \,\square^{n-l}R\;,
    \\
    {\Omega}_2{{}_{ab}} &=\sum_{n=1}^\infty f_{2,n}\sum_{l=0}^{n-1}{\nabla_a} \square^l {S^{cd}}\,{\nabla_b} \square^{n-l-1}{S_{cd}}\;, & \Theta_2 &=\sum_{n=1}^\infty f_{2,n}\sum_{l=0}^{n-1}\square^l {S^{cd}} \,\square^{n-l}{S_{cd}}\;,
    \\
    {\Omega}_3{{}_{ab}} &=\sum_{n=1}^\infty f_{3,n}\sum_{l=0}^{n-1}{\nabla_a} \square^l {C^{cdef}}\,{\nabla_b} \square^{n-l-1}{C_{cdef}}\;, & \Theta_3 &=\sum_{n=1}^\infty f_{3,n}\sum_{l=0}^{n-1}\square^l {C^{cdef}} \,\square^{n-l}{C_{cdef}}\;,
\end{aligned}
\\
\begin{aligned}
    {\Upsilon}_2{{}_{ab}} &=\sum_{n=1}^\infty f_{2,n}\sum_{l=0}^{n-1} {\nabla_c}\big[\square^l {S^{cd}}\,{\nabla_{(a}} \square^{n-l-1}{S_{{b)}d}}-{\nabla_{(a}}\square^l {S^{cd}} \,\square^{n-l-1}{S_{{b)}d}}\big]\;,
    \\
    {\Upsilon}_3{{}_{ab}} &=\sum_{n=1}^\infty f_{3,n}\sum_{l=0}^{n-1} {\nabla_c}\big[\square^l {C^{cdef}}\,{\nabla_{(a}} \square^{n-l-1}{C_{{b)}def}}-{\nabla_{(a}}\square^l {C^{cdef}}\, \square^{n-l-1}{C_{{b)}def}}\big]\;.
\end{aligned}
\end{gathered}
\end{equation}
The presence of the non-local non-linear expressions such as $\bs{\Omega}_i$, $\bs{\Upsilon}_i$, and ${\Theta_i}$ in the field equations makes any attempt of finding exact solutions extremely challenging. Not only are the expressions very large and convoluted, but there exist (almost) no mathematical methods for solving such non-local non-linear equations. Thus, it is reasonable to focus the attention first on the class of geometries that reduce the field equations either to i) \textit{local} non-linear equations, or to ii) non-local \textit{linear} equations.

The known exact solutions of the first type are, for instance, the bouncing cosmologies \cite{Biswas:2005qr,Biswas:2010zk,Koshelev:2012qn,Biswas:2012bp,Koshelev:2017tvv,Koshelev:2020xby,Koshelev:2020foq}, which assume the ansatz ${\square R=r_1 R+r_2}$ with two constants ${r_1}$ and ${r_2}$. The solutions generated by this or similar recurrent formulas with curvature typically depend only on the values and derivatives of the form-factors $\mathcal{F}_i$ at a few specific points (such as $r_1$). The known exact solutions of the second type are, for example, the gravitational waves found in \cite{Kilicarslan:2019njc,Dengiz:2020xbu}. These solutions are interesting for their strong dependence on all values of the non-local form-factors $\mathcal{F}_i$, which is a typical feature that we see in all solutions in the weak-field regime of IDG. In this sense, the second type of ansatz represents a rather unique balance between mathematical difficulty and physical significance due to \textit{linearity} and \textit{non-locality} of the resulting field equations.

In this paper we focus on this second type of geometry ansatz by investigating the class of \textit{almost universal spacetimes}, which has been found recently in \cite{Kuchynka:2018ezs}. As we will see, this class fits perfectly our demand for reducing the full IDG field equations to a single non-local but linear equation. Thus, it offers a great opportunity to find many exact physically interesting solutions. Almost universal spacetimes generalize universal spacetimes studied in \cite{Coley:2008th,Hervik:2013cla,Hervik:2015mja,Hervik:2017sdr}. These spacetimes have a special property that any correction to the Einstein--Hilbert action of GR constructed from curvature invariants does not contribute to the field equations (up to the modification of the cosmological constant). Relaxing appropriately the conditions imposed on the metric to be universal, one obtains the class of almost universal spacetimes for which the corrections do not vanish anymore, however, the resulting field equations are linear.

The paper is organized as follows: In Section \ref{sc:AUS} we review the class of almost universal spacetimes and examine their most important properties. In Section \ref{sc:RFE} we show that this class of metrics simplifies the field equations of IDG to the extent that we can solve them by means of eigenfunction expansion or using the heat kernel method. In Section \ref{sc:EWN0}, we employ these techniques to find exact gravitational-wave solutions generated by null sources, i.e., non-local versions of the Hotta--Tanaka (${\Lambda\neq0}$) and Aichelburg--Sexl (${\Lambda=0}$). In Section \ref{sc:CU}, we compute components of the curvature tensors of these solutions in a natural frame. The paper is concluded with a brief discussion of our results in Section \ref{sc:C}. Appendices \ref{app:SCid}, \ref{app:NP}, \ref{app:ppframe}, and \ref{app:tnskundttaunz} provide supplementary material.

\section{Almost universal spacetimes}\label{sc:AUS}

The so-called \textit{universal metrics} can be employed as a useful and simple method for finding exact vacuum solutions of modified theories of gravity with Lagrangian being an analytic function of the metric, the Riemann tensor, and its covariant derivatives of an arbitrary order,
\begin{equation}
    L = L(\bs{g},\bs{R}, \bs{\nabla R}, ...)\;.
\end{equation}
Universal metrics are defined in the way that any rank-2 tensor corresponding to the field equations of any such theory is proportional to the metric. This immediately implies that universal metrics are Einstein and the field equations reduce to just one algebraic constraint relating the constant Ricci scalar with parameters of the theory. In the case of IDG \eqref{eq:action}, this approach leads to solutions equivalent to GR solutions which do not exhibit non-local effects and therefore are not of interest to this study.

However, the class of universal metrics can be appropriately generalized by relaxing the condition on the rank-2 tensors.
The \textit{almost universal spacetimes} \cite{Kuchynka:2018ezs} (or equivalently the \textit{TN spacetimes}\footnote{TN stands for the traceless part of $\bs{B}$ being of type N.}) are thus defined as spacetimes, for which every symmetric rank-2 tensor $\bs{B}$ constructed from the metric, the Riemann tensor, and its covariant derivatives of an arbitrary order takes the form
\begin{equation}\label{eq:TN}
    B_{ab}  = \lambda\, g_{ab} + \phi\, l_a l_b\;,
\end{equation}
where $\lambda$ is a constant, $\phi$ is a scalar function, and $\bs{l}$ is a null vector. Note that different tensors $\bs{B}$ generally give rise to different $\lambda$ and $\phi$. Obviously, the Ricci tensor of a TN spacetime is of the form \eqref{eq:TN} and therefore the Ricci scalar is constant and the traceless Ricci tensor $\bs{S}$ is of type N, i.e., it has only one component,
\begin{equation}\label{eq:TFRicci}
    S_{ab} = 2 \Phi_{22}\,l_a l_b\;.
\end{equation}
TN spacetimes are necessarily algebraically special Kundt spacetimes (which will be discussed below) with scalar curvature invariants being constant. Conversely, all the Kundt spacetimes of the Weyl type III or N and traceless Ricci type N are TN spacetimes.

In general, both the traceless Ricci tensor $\bs{S}$ and the Weyl tensor $\bs{C}$ contribute to $\phi$ in \eqref{eq:TN}, however, it turns out that if the Weyl tensor of a Weyl type III TN spacetime obeys\footnote{Note that, in higher dimensions, the Weyl tensor must also satisfy another condition ${C_{acde} C_b{}^{cde} = 0}$, which is met in four dimensions due to the well-known identity ${C_{acde} C^{bcde} = \frac{1}{4} \delta_a{}^b C_{cdef} C^{cdef}}$.}
\begin{equation}\label{eq:F2}
    C^\text{II}_{ab} \equiv \nabla_a C_{cdef} \nabla_b C^{cdef} = 0\;,
\end{equation}
then the last term in \eqref{eq:TN} reduces to
\begin{equation}\label{eq:TNS}
    \phi\, l_a l_b = \mathcal{B}(\square) S_{ab}\;,
\end{equation}
where $\mathcal{B}(\square)$ is an analytic operator with constant coefficients $b_k$,
\begin{equation}
    \mathcal{B}(\square)=\sum_{k=0}^{\infty} b_k \square^{k}\;.
\end{equation}
This special subclass of TN spacetimes is referred to as the \textit{TNS spacetimes}. One of the key points in the proof of \eqref{eq:TNS} is the statement that $\bs{\nabla}^k\bs{S}$ and $\bs{\nabla}^k\bs{C}$, ${k\geq0}$, contribute to rank-2 tensors $\bs{B}$ only linearly and their mixed terms vanish. Using these properties of $\bs{S}$ and $\bs{C}$, we are able to derive the following useful identities (see Appendix~\ref{app:SCid}):
\begin{equation}\label{eq:SCid}
\begin{aligned}
\nabla_c\nabla_b\square^n S^{ca} &=\frac13 R\,\square^n S_b^a\;,
\\
\nabla_c\nabla_d\square^n C_b{}^{cda} &=-\frac12\bigg(\square+\frac{R}{3}\bigg)^{\!\!n}\bigg(\square-\frac{R}{3}\bigg)S^a_b\;.
\end{aligned}
\end{equation}

As mentioned above, TN spacetimes are very closely related to the Kundt spacetimes, which is one of the most important families containing exact solutions of GR and modified theories of gravity. Note that, for instance, the well known pp-waves and other non-expanding gravitational waves of various kinds belong to this class. The \textit{Kundt spacetimes} are geometrically defined as geometries admitting a congruence of null geodesics which is non-expanding and has zero shear and twist. If the null vector $\bs{l}$ is tangent to such congruence then the geodesic, expansion-free, shear-free, and twist-free condition read
\begin{equation}\label{eq:Kundtconds}
    l^b \nabla_b l^a = 0\;,
    \quad
    \theta \equiv \frac{1}{2} \nabla_a l^a = 0\;,
    \quad
    \sigma\bar\sigma \equiv \frac{1}{2} \nabla_{(a} l_{b)} \nabla^a l^b - \theta^2 = 0\;,
    \quad
    \omega^2 \equiv \frac{1}{2} \nabla_{[a}l_{b]} \nabla^a l^b = 0\;,
\end{equation}
respectively. Starting with these assumptions on the null congruence $\bs{l}$, one can obtain a general form of the Kundt metric\footnote{The symbol ${\vee}$ denotes a symmetric tensor product of two tensors of the same type, ${\bs{a}\vee\bs{b}=\bs{a}\bs{b}+\bs{b}\bs{a}}$. The inverse metric reads
\begin{equation*}
    \bs{g}^{-1}=\big[{-}\bs{\pp}_{u}+(H+P^2W\bar{W})\bs{\pp}_{r}-P^2\bar{W}\bs{\pp}_{\zeta}-P^2W\bs{\pp}_{\bar\zeta}\big]\vee\bs{\pp}_{r}+P^2\bs{\pp}_{\zeta}\vee\bs{\pp}_{\bar\zeta}\;.
\end{equation*}} \cite{Kundt1961} (see also \cite{Stephani:2003tm,Griffiths:2009dfa})
\begin{equation}\label{eq:Kundt}
    \bs{g} = -\bs{\dd}u\vee(H \bs{\dd}u+\bs{\dd}r + W \bs{\dd}\zeta + \bar{W} \bs{\dd}\bar\zeta) + P^{-2} \bs{\dd}\zeta \vee\bs{\dd}\bar\zeta\;,
\end{equation}
where $P(u,\zeta,\bar\zeta)$ and $H(u,r,\zeta,\bar\zeta)$ are real functions, and $W(u,r,\zeta,\bar\zeta)$ is a complex function. The coordinate $r$ is an affine parameter along the null geodesics $\bs{l} = \bs{\partial}_r$ corresponding to non-expanding rays along which gravitational waves propagate. For a given value of $r$, the 2-dimensional surfaces ${u = \text{const.}}$ are submanifolds ${\mathbb{M}\subset M}$ equipped with the Riemannian metrics
\begin{equation}\label{eq:2dmetric}
    \bs{q}=P^{-2} \bs{\dd}\zeta \vee\bs{\dd}\bar\zeta\;,
\end{equation}
which characterize geometries of the wave surfaces. 

Weyl type III and N Kundt spacetimes were studied in the context of GR with a null matter content ${T_{ab} \propto l_a l_b}$ in \cite{Griffiths:2003bk} (see also \cite{Griffiths:2009dfa}). Since their traceless Ricci tensor is of type N (where $R_{ab}$ takes the form \eqref{eq:TN} where ${\lambda = R/4}$ and $\phi$ corresponds to the null radiation term), these geometries contain all Weyl type III/N TN spacetimes. The conditions imposed on the components of the Ricci and Weyl tensor have several implications: The Riemannian 2-space ${(\mathbb{M},\bs{q})}$ is of constant curvature and the function $P$ can be put to the form
\begin{equation}\label{eq:TNKundtP}
    P = 1 + \frac{R}{24} \zeta \bar\zeta
\end{equation}
using coordinate freedom. The spin coefficient $\tau$ (see Appendices \ref{app:NP} and \ref{app:ppframe}) can be integrated out to get
\begin{equation}\label{eq:TNKundttauQ}
    \tau = \frac{1}{Q}\bigg({-}b + \frac{R}{12} a \zeta + \frac{R}{24} \bar{b}\zeta^2\bigg)\;, 
    \quad Q = a + \bar{b}\zeta + b \bar\zeta - \frac{R}{24} a \zeta\bar\zeta\;,
\end{equation}
with $a$ and $b$ being real and complex constants, respectively. Finally, it turns out that
\begin{equation}\label{eq:TNKundtWH}
    W = \frac{2\bar\tau}{P} r + W^\circ\;, \quad H = -\left(\tau\bar\tau + \frac{R}{24} \right)r^2 + 2 G^\circ r + H^\circ\;,
\end{equation}
where ${H^\circ = H^\circ(u, \zeta, \bar\zeta)}$, ${W^\circ = W^\circ(u, \zeta, \bar\zeta)}$ are arbitrary functions independent of $r$. Function $G^\circ$ is then determined by
\begin{equation}\label{eq:TNKundtG}
    G^\circ = -\frac12 P(\tau W^\circ + \bar\tau W^\circ) - \frac{R}{24} \int W^\circ \mathrm{d}\zeta\;.
\end{equation}
Consequently, the family of Kundt metrics \eqref{eq:Kundt} with \eqref{eq:TNKundtP}--\eqref{eq:TNKundtG} is equivalent to the class of all Weyl type III/N TN spacetimes. Furthermore, such spacetimes are TNS if the Weyl tensor satisfies \eqref{eq:F2}. This additional condition in the language of NP formalism reads
\begin{equation}\label{eq:F2_NP}
    C^\text{II}_{ab}=96(\tau^2 \Psi_3^2 + \bar\tau^2 \bar\Psi_3^2) l_a l_b=0\;.
\end{equation}

In a parallel-propagated frame subject to \eqref{eq:taucond} (see Appendix \ref{app:ppframe} for an example of such a frame),
one can write the action of $\square^k$ on $S_{ab}$ \eqref{eq:TFRicci} as 
\begin{equation}\label{eq:boxk}
    \square^k S^a_b=2 l^a l_b(\mathcal{D}{+}R/2)^k \Phi_{22}\;,
\end{equation}
where the linear differential operator $\mathcal{D}$ is defined as
\begin{equation}\label{eq:mathcalD}
    \mathcal{D} \equiv \square + 4 (\tau \bar\delta + \bar\tau \delta) + 4 \tau\bar\tau\;.
\end{equation}
Its action on a scalar function ${f=f(u,\zeta,\bar{\zeta})}$ independent of the coordinate $r$ is given by
\begin{equation}
    \mathcal{D}f = \triangle f + 2P \tau \pp_{\zeta}f + 2P \bar{\tau} \pp_{\bar{\zeta}}f + 4\bar{\tau}\tau f\;,
\end{equation}
where we introduced Laplace operator $\triangle$ associated with the homogeneous metric ${\bs{q}}$ on $\mathbb{M}$,
\begin{equation}
    \triangle \equiv 2P^2\pp_{\zeta}\pp_{\bar\zeta}\;.
\end{equation}

In the generic case ${\tau \neq 0}$, the function
${W^\circ = W^\circ(u, \zeta)}$ is independent of $\bar\zeta$ and the components $\Psi_3$ and $\Phi_{22}$ are given by
\begin{equation}\label{eq:phi22}
    \Psi_3 = - \frac{P^2}{Q}(Q\tau W^\circ)_{,\zeta}\;, \quad \Phi_{22} = \frac12 \bigg(\mathcal{D} H^\circ + \frac{R}{3} H^\circ + V\bigg)\;,
\end{equation}
where we denoted
\begin{equation}\label{eq:Ntaunonzero}
    V \equiv 2P\tau \bar{W}^\circ (P^2 W^\circ)_{,\zeta} +  2P\bar\tau W^\circ (P^2 \bar{W}^\circ)_{,\bar\zeta} + 2P^2(2 \tau\bar\tau + R/4) W^\circ \bar{W}^\circ\;.
\end{equation}
It also turns out that $\mathcal{D}f$ can be written in a more compact and convenient form,
\begin{equation}\label{eq:mathcalDonf}
    \mathcal{D}f =  \frac{Q}{P} \triangle\left(\frac{P}{Q}f\right) - \frac{R}{6} f\;,
\end{equation}
containing $\triangle$ as the only differential operator.

On the other hand, in the special case ${\tau=0}$, which is possible only for ${R=0}$, the function ${W^\circ = W^\circ(u,\bar\zeta)}$ is independent of $\zeta$ and the operator $\mathcal{D}$ reduces to Laplace operator,
\begin{equation}\label{eq:DfDeltaf}
    \mathcal{D}f=\triangle f\;.
\end{equation}
The components $\Psi_3$ and $\Phi_{22}$ read
\begin{equation}\label{eq:phi22tau0}
    \Psi_3 = \frac12 \bar{W}^\circ_{,\zeta\zeta}\;, \quad    \Phi_{22} = \frac12 \left(\mathcal{D} H^\circ + V \right)\;,
\end{equation}
where $V$ is now defined by
\begin{equation}\label{eq:Nzerotaulambda}
    V \equiv - W^\circ W^\circ_{,\bar\zeta\bar\zeta} - \bar{W}^\circ \bar{W}^\circ_{,\zeta\zeta} - W^\circ_{,u\bar\zeta} - \bar{W}^\circ_{,u\zeta} - W^\circ_{,\bar\zeta}{}^2 - \bar{W}^\circ_{,\zeta}{}^2\;.
\end{equation}


\section{Reduction of field equations}\label{sc:RFE}

Let us now apply the properties of almost universal spacetimes from the previous section (specifically Weyl type III/N TNS) to the field equations of IDG \eqref{eq:fieldeq}. Since $R$ is constant, and all quadratic and mixed terms of $\bs{\nabla}^k\bs{S}$ and $\bs{\nabla}^k\bs{C}$, ${k\geq 0}$ vanish, many expressions including all $\bs{\Omega}_i$, ${\Upsilon_i}$, and $\Theta_i$ in \eqref{eq:LK} vanish. As a consequence, the field equations \eqref{eq:fieldeq} drastically simplify. We obtain
\begin{equation}
\begin{aligned}
E^a_b &=S^a_b{-}\frac14 R\delta^a_b{+}\Lambda\delta^a_b +2\mathcal{F}_1(0) RS^a_b+\frac{1}{2}R\mathcal{F}_2(\square) S^a_b
\\
&\feq-2\nabla_c\nabla_b\mathcal{F}_2(\square) S^{ca} +\square\mathcal{F}_2(\square) S^a_b+\delta^a_b \nabla_c \nabla_d \mathcal{F}_2(\square) S^{cd}-4\nabla_c\nabla_d\mathcal{F}_3(\square) C_b{}^{cda}\;,
\end{aligned}
\end{equation}
where only the terms linear in curvature tensors $\bs{S}$ and $\bs{C}$ survived. Furthermore, using the identities \eqref{eq:SCid} we can rewrite the field equations in the form that contains only $\delta^a_b$ and $\square^k S^a_b$,
\begin{equation}\label{eq:fieldeqSdel}
E^a_b =(\Lambda{-}R/4)\delta^a_b+\big[1+2\mathcal{F}_1(0) R+\mathcal{F}_2(\square)(\square{-}R/6)+2\mathcal{F}_3(\square{+}R/3)(\square{-}R/3)\big] S^a_b\;,
\end{equation}
which is clearly of the TNS form \eqref{eq:TN} with \eqref{eq:TNS}.

Let us consider the parallel-propagated frame from Appendix \ref{app:ppframe}, which allows us to use the formula \eqref{eq:boxk} for the action of $\square^k$ on $S^a_b$ in terms of the operator ${\mathcal{D}}$ from \eqref{eq:mathcalD}. Hence, we can write
\begin{equation}\label{eq:trEtrfE}
    E = 4\Lambda{-}R\;,
    \quad
    \tilde{E}^a_b = 2l^a l_b \big[1+2\mathcal{F}_1(0)R+\mathcal{F}_2(\mathcal{D}{+}R/2)(\mathcal{D}{+}R/3)+2\mathcal{F}_3(\mathcal{D}{+}5R/6)(\mathcal{D}{+}R/6)\big]\Phi_{22}\;,
\end{equation}
where we split the field equations \eqref{eq:fieldeqSdel} in the trace part $E\equiv E^a_a$ and trace-free part ${\tilde{E}^a_b\equiv E^a_b-\frac14 E\delta^a_b}$. Due to the structure of the field equations, we may study the solutions admitting null radiation described by the traceless energy-momentum tensor ${T^a_b=J\, l^a l_b}$, where $J$ is an arbitrary scalar function. The first equation of \eqref{eq:trEtrfE} implies ${R=4\Lambda}$ while the second one leads to
\begin{equation}\label{eq:Gphi22J}
    2\mathcal{G}(\mathcal{D})\Phi_{22}=J\;,
\end{equation}
where we introduced an analytic operator ${\mathcal{G}(\mathcal{D})}$,
\begin{equation}
\mathcal{G}(\mathcal{D})\equiv 1+8\mathcal{F}_1(0)\Lambda+\mathcal{F}_2(\mathcal{D}{+}2\Lambda)(\mathcal{D}{+}4\Lambda/3)+2\mathcal{F}_3(\mathcal{D}{+}10\Lambda/3)(\mathcal{D}{+}2\Lambda/3)\;.
\end{equation}
To emphasize the dependence on the length scale of non-locality $\ell$, we can write the explicit expression as
\begin{equation}
   \mathcal{G}(\mathcal{D})= 1+8 \hat{f}_{1,0}\ell ^2+ \sum_{n=0}^{\infty } \left[\hat{f}_{2,n}\left(\mathcal{D}+4 \Lambda /3\right) (\mathcal{D}+2 \Lambda )^n+2  \hat{f}_{3,n}\left(\mathcal{D}+2\Lambda/3\right) \left(\mathcal{D}+10\Lambda/3\right)^n\right]\ell^{2 n+2}\;.
\end{equation}
Inserting the expression for ${\Phi_{22}}$ from \eqref{eq:phi22} (and \eqref{eq:phi22tau0} for ${\tau=0}$) into \eqref{eq:Gphi22J}, we arrive at the equation
\begin{equation}\label{eq:redfieldeq}
    \mathcal{G}(\mathcal{D})\big[(\mathcal{D}+4\Lambda/3)H^\circ+V\big]=J\;.
\end{equation}
This equation contains two unknown functions, $H^{\circ}$ and $W^{\circ}$, which enters via $V$. As it is discussed in Appendix~\ref{app:tnskundttaunz}, the function $W^{\circ}$ is not completely arbitrary for ${\tau\neq0}$ because one has to ensure that the condition \eqref{eq:F2_NP}, ${C^\text{II}_{ab}=0}$, is satisfied.

For simplicity (and for comparison with the GR solutions studied in the literature), we fix the function $W^{\circ}$ by an additional condition
\begin{equation}\label{eq:V0}
     V=0\;.
\end{equation}
In the generic situation ${\tau\neq 0}$, the choice \eqref{eq:V0} leads immediately to ${W^{\circ}=0}$ (Weyl type specializes to N) because the function ${V}$ from \eqref{eq:Ntaunonzero} for any ${W^{\circ}\neq0}$ can be put to the form
\begin{equation}
    V=\frac{2P^3 W^\circ \bar{W}^\circ}{Q}  \left( \tau Q(\ln W^\circ)_{,\zeta} + \bar\tau Q(\ln \bar{W}^\circ)_{,\bar\zeta} + \frac{Q}{P}(2\tau\bar\tau + 2 \tau P_{,\zeta} + 2 \bar\tau P_{,\bar\zeta} + R/4) \right)\;.
\end{equation}
However, such an expression can never vanish, because the first term in the parentheses is independent of $\bar\zeta$, the second term is independent of $\zeta$, while $\zeta$ and $\bar\zeta$ are mixed in the third term. The special case, ${\tau=\Lambda=0}$, provides more options for $W^{\circ}$ satisfying \eqref{eq:V0}. From \eqref{eq:Nzerotaulambda}, we can see that $V$ vanishes if ${W^\circ W^\circ_{,\bar\zeta} + W^\circ_{,u} = z(u)}$ with an arbitrary function $z(u)$. This equation has a general solution that is given implicitly by (see \cite{Polyanin:2001})
\begin{equation}\label{eq:WcircVzerotauzero}
    \bar\zeta = (W^\circ - Z(u))u + H(W^\circ - Z(u)) +  \int du\, Z(u)\;,
    \quad
    Z(u) \equiv\int du\, z(u)\;,
\end{equation}
where $H$ is an arbitrary functions. Since the conditions \eqref{eq:V0} implies that either ${W^{\circ}=0}$ or ${\tau=0}$, the corresponding geometries automatically satisfy condition \eqref{eq:F2_NP}, ${C^\text{II}_{ab}=0}$.

Let us further simplify the field equation \eqref{eq:redfieldeq} (with \eqref{eq:V0}) using other properties of TNS spacetimes. For ${\tau\neq 0}$, the repeated use of the identity \eqref{eq:mathcalDonf} leads to the relation
\begin{equation}
    \mathcal{D}^n f=\frac{Q}{P}\left( \triangle-\frac23\Lambda\right)^{\!\!n} \left(\frac{P}{Q}f\right)\;,
\end{equation}
which allows us to rewrite the operator $\mathcal{G}(\mathcal{D})$ as a function of the Laplace operator $\triangle$ sandwiched between the factors ${P/Q}$ and ${Q/P}$,
\begin{equation}
    \mathcal{G}(\mathcal{D})f=\frac{Q}{P}\mathcal{G}(\triangle-2\Lambda/3)\left(\frac{P}{Q}f\right)\;.
\end{equation}
As a result, the field equation can now be recast in a much more tractable form
\begin{equation}\label{eq:redfielddeltatilde}
    \mathcal{G}(\triangle-2\Lambda/3)\left(\triangle+2\Lambda/3\right)\hat{H}^{\circ}=\hat{J}\;,
\end{equation}
where we introduced the re-scaled quantities
\begin{equation}
    \hat{H}^{\circ}\equiv\frac{P}{Q}H^{\circ}\;,
    \quad
    \hat{J}\equiv\frac{P}{Q}J\;.
\end{equation}
In the special case ${\tau=\Lambda=0}$, no re-scaling of $H^{\circ}$ and $J$ is needed thanks to \eqref{eq:DfDeltaf}; the field equation reads
\begin{equation}\label{eq:redfielddeltaspecial}
    \mathcal{G}(\triangle)\triangle H^{\circ}=J\;.
\end{equation}

The fact that we arrived to the equations \eqref{eq:redfielddeltatilde} and \eqref{eq:redfielddeltaspecial} containing a dependence on a single operator $\triangle$ is of a great significance. Since $\triangle$ is the Laplace operator on 2-dimensional spaces of constant curvature, the problem is always solvable by known mathematical techniques, which is an important consequence of the class of TNS spacetimes. Specifically, the equation \eqref{eq:redfielddeltatilde}\footnote{Here we focus on the generic case ${\tau\neq 0}$, however, the discussion for ${\tau=\Lambda=0}$ is very similar.} is a linear partial differential equation for a scalar function ${\hat{H}^{\circ}=\hat{H}^{\circ}(u,\mathrm{x})}$, ${\mathrm{x}\in \mathbb{M}}$. Such an equation can be solved using expansion in the eigenfunctions ${\psi_{\alpha}=\psi_{\alpha}(\mathrm{x})}$ of ${\triangle}$,
\begin{equation}\label{eq:eigenvalueproblem}
    \triangle\psi_{\alpha}=-\mu_{\alpha}^2\psi_{\alpha}\;,
\end{equation}
where $\mu_{\alpha}^2$ denotes the corresponding eigenvalues. Thus, our aim is to find solutions in the space of functions that are linear combinations of $\psi_{\alpha}$. For this purpose we decompose ${\hat{J}}$ in the eigenfunctions,
\begin{equation}\label{eq:expJN}
    \hat{J}(u,\mathrm{x})= \sumint_{\alpha}\hat{J}_{\alpha}(u)\psi_{\alpha}(\mathrm{x})\;,
\end{equation}
where $\sumint_{\alpha}$ denotes summation $\sum_{\alpha}$ if the spectrum is discrete and integration $\int_{\alpha}$ if the spectrum is continuous.

A general solution (expressible using $\psi_{\alpha}$) splits into a general homogeneous solution plus a single particular solution,
\begin{equation}
    \hat{H}^{\circ}=\hat{H}^{\circ}_{\textrm{hom}}+\hat{H}^{\circ}_{\textrm{part}}\;,
\end{equation}
where $\hat{H}^{\circ}_{\textrm{hom}}$ satisfies homogeneous equation everywhere in $\mathbb{M}$ (i.e., singular functions are excluded),
\begin{equation}\label{eq:homeq}
    \mathcal{G}(\triangle-2\Lambda/3)(\triangle+2\Lambda/3)\hat{H}^\circ_{\textrm{hom}}=0\;,
\end{equation}
while $\hat{H}^{\circ}_{\textrm{part}}$ solve the equation with a given source ${J}$, respectively. The number of independent solutions $\hat{H}^\circ$ is given purely by the homogeneous parts $\hat{H}^\circ_{\textrm{hom}}$. Therefore, it depends on the choice of the form-factor $\mathcal{G}$. In what follows, we assume that $\mathcal{G}$ has no zeros in the complex plane which guarantees that the number of solutions is the same as in the local case ${\ell=0}$ (i.e., ${\mathcal{G}=1}$). Consequently, the space of solutions of \eqref{eq:homeq} is then equivalent to the space of solutions of
\begin{equation}
    (\triangle+2\Lambda/3)\hat{H}^{\circ}_{\textrm{hom}}=0\;.
\end{equation}
According to the Weierstrass factorization theorem \cite{greene2006function} an entire function $\mathcal{G}$ with no zeros in the complex plane can be written as exponential of an entire function $\mathcal{A}$. Thus, we can write\footnote{Although, we do not study ghost instabilities in this paper, it should be pointed out that \eqref{eq:GA} is compatible with the ghost-free condition in (anti-)de Sitter background \cite{Biswas:2016egy}. Specifically, \eqref{eq:GA} arises, for example, in the following choice of form-factors $\mathcal{F}_i$ (explicitly mentioned in \cite{SravanKumar:2019eqt}):
\begin{equation*}
    \mathcal{F}_1(\square)=\mathcal{F}_2(\square)=0\;,\quad \mathcal{F}_3(\square)=\frac{1}{2}\frac{e^{-\mathcal{A}(\square-2R/3)}-1}{\square-2R/3}\;.
\end{equation*}
}
\begin{equation}\label{eq:GA}
    \mathcal{G}(\triangle-2\Lambda/3)=e^{-\mathcal{A}(\triangle)}\;, 
\end{equation}
To find a particular solution of \eqref{eq:redfielddeltatilde} with \eqref{eq:GA}, we can formally invert the equation~\eqref{eq:redfielddeltatilde} with the help of \textit{eigenfunction expansion}. After expanding $\hat{H}^{\circ}_{\textrm{part}}$ and the source $\hat{J}$ (see \eqref{eq:expJN}) in terms of ${\psi_{\alpha}}$ satisfying \eqref{eq:eigenvalueproblem}, we can write the particular solution as
\begin{equation}\label{eq:Hzerosol}
    \hat{H}^{\circ}_{\textrm{part}} =\sumint_\alpha\frac{e^{\mathcal{A}(-\mu_{\alpha}^2)}}{-\mu_{\alpha}^2+2\Lambda/3}\hat{J}_{\alpha}\psi_{\alpha}=\frac{e^{\mathcal{A}(\triangle)}}{\triangle+2\Lambda/3}\hat{J} \;.
\end{equation}

As a simple example of $\mathcal{A}$, we will consider
\begin{equation}\label{eq:Az}
     \mathcal{A}(\triangle)=\ell^2 \triangle\;.
\end{equation}
This choice is also important because it allows us to find the solution for ${\hat{H}^{\circ}_{\textrm{part}}}$ alternatively using the \textit{heat kernel method} (see, e.g., \cite{Frolov:2015bia}). To understand this method let us start with the heat equation for an auxiliary scalar field ${\phi=\phi(s,\mathrm{x})}$ on the manifold $\mathbb{M}$ with the metric $\bs{q}$ with a new parameter ${s\geq0}$,
\begin{equation}\label{eq:heat}
    \triangle \phi=\pp_s \phi\;,\quad \phi|_{s=0}=f\;.
\end{equation}
Here, ${f=f(\mathrm{x})}$ is an arbitrary function on $\mathbb{M}$ representing the initial condition at ${s=0}$. The corresponding solution ${\phi}$ is then given by the heat kernel ${K=K(s;\mathrm{x},\mathrm{x}')}$,
\begin{equation}\label{eq:phisx}
    \phi(s,\mathrm{x}) =\!\!\int\limits_{\mathrm{x}'\in \mathbb{M}}\!\!\!\mathfrak{q}^{\frac12}(\mathrm{x}')K(s;\mathrm{x},\mathrm{x}')f(\mathrm{x}')\;.
\end{equation}
Another useful characterization of $\phi$ can be obtained by expanding $\phi$ in terms of eigenfunctions ${\psi_{\alpha}}$ satisfying \eqref{eq:eigenvalueproblem}. The heat equation \eqref{eq:heat} is then formally solved by
\begin{equation}\label{eq:phisx2}
    \phi(s,\mathrm{x})=\sumint_{\alpha}e^{-s\mu_{\alpha}^2}f_{\alpha}(\mathrm{x})\psi_{\alpha}(\mathrm{x})=e^{s\triangle}f(\mathrm{x})\;.
\end{equation}
Comparing \eqref{eq:phisx2} with \eqref{eq:phisx}, we get an important integral representation of the exponential operator $e^{s\triangle}$,
\begin{equation}\label{eq:expstriangle}
    e^{s\triangle}f(\mathrm{x})=\!\!\int\limits_{\mathrm{x}'\in \mathbb{M}}\!\!\!\mathfrak{q}^{\frac12}(\mathrm{x}')K(s;\mathrm{x},\mathrm{x}')f(\mathrm{x}')\;.
\end{equation}
Returning to the particular solution ${\hat{H}^{\circ}_{\textrm{part}}}$, we can now rewrite \eqref{eq:Hzerosol} with \eqref{eq:Az} by means of \eqref{eq:expstriangle},
\begin{equation}\label{eq:solheatker}
\begin{aligned}
    \hat{H}^{\circ}_{\textrm{part}}(u,\mathrm{x})&=\frac{e^{\ell^2 \triangle}}{\triangle+2\Lambda/3}\hat{J}(u,\mathrm{x})
    =-e^{-2\Lambda\ell^2/3}\int_{\ell^2}^{\infty} \!ds \,e^{s (\triangle+2\Lambda/3)}\hat{J}(u,\mathrm{x})
    \\
    &=-e^{-2\ell^2\Lambda/3}\int_{\ell^2}^{\infty} \!ds \,e^{2s\Lambda/3}\!\!\!\int\limits_{\mathrm{x}'\in \mathbb{M}} \!\!\mathfrak{q}^{\frac12}(\mathrm{x}')K(s,\mathrm{x},\mathrm{x}')\hat{J}(u,\mathrm{x}')\;,
\end{aligned}
\end{equation}
provided that the integral converges. This trick is especially useful for ${\Lambda<0}$ because the heat kernel is well known on the 2-space of constant negative curvature while the eigenfunction expansion is rather difficult here.


\section{Examples}\label{sc:EWN0}
In this section, we provide some explicit solutions for all three choices of $\Lambda$ corresponding to three options for Gaussian curvature of the 2-space ${(\mathbb{M},\bs{q})}$. As the homogeneous solutions ${H_{\textrm{hom}}^{\circ}}$ are not affected by the non-locality, we focus on finding the particular solutions ${H_{\textrm{part}}^{\circ}}$ only.

\subsection{Solutions with ${\Lambda>0}$}
For a positive cosmological constant $\Lambda$, the 2-space corresponds to a sphere $\mathbb{S}^2$ with the standard homogeneous metric $\bs{q_+}$ of positive Gaussian curvature ${\Lambda/3>0}$. Using the coordinate transformation to the \textit{spherical coordinates}\footnote{The Jacobian matrix of \eqref{eq:transzetap} is
\begin{equation*}
    \begin{bmatrix}
    \pp_{\zeta}\chi & \pp_{\bar{\zeta}}\chi
    \\
    \pp_{\zeta}\varphi & \pp_{\bar{\zeta}}\varphi
    \end{bmatrix} = 
    \sqrt{\frac{\Lambda}{6}}\begin{bmatrix}
    \cos^2\left(\frac{\chi}{2}\right)e^{-i\varphi} & \cos^2\left(\frac{\chi}{2}\right)e^{i\varphi}
    \\
    -\frac{i}{2}\cot\left(\frac{\chi}{2}\right)e^{-i\varphi} & \frac{i}{2}\cot\left(\frac{\chi}{2}\right)e^{i\varphi}
    \end{bmatrix}\;.
\end{equation*}}
\begin{equation}\label{eq:transzetap}
    \zeta=\sqrt{\frac{6}{\Lambda}} \tan \left(\frac{\chi}{2}\right) e^{i \varphi}\;,
    \quad
    \chi\in(0,\pi)\;, \quad \varphi\in(0,2\pi)\;,
\end{equation}
we can bring the metric \eqref{eq:2dmetric} to a more familiar form,
\begin{equation}\label{eq:qplus}
    \bs{q}_+=\frac{3}{\Lambda}(\bs{\dd} \chi^2+\sin^2\chi\bs{\dd} \varphi^2)\;.
\end{equation}
The Laplace operator $\triangle$ in these coordinates is given by
\begin{equation}\label{eq:laplpos}
    \triangle f =\frac{\Lambda}{3}\frac{1}{\sin\chi}\pp_{\chi}\big(\sin\chi\pp_{\chi}f\big)+\frac{\Lambda}{3}\frac{1}{\sin^2\chi}\pp_{\varphi}^2 f
\end{equation}
and the functions ${P}$ and ${Q}$ from \eqref{eq:TNKundtP} and \eqref{eq:TNKundttauQ} read
\begin{equation}\label{eq:PQpos}
    P=\frac{1}{\cos^2\left(\frac{\chi}{2}\right)}\;,
    \quad
    Q=a\frac{\cos{\chi}}{\cos^2\left(\frac{\chi}{2}\right)}+\sqrt{\frac{6}{\Lambda}}\big(\bar{b}e^{i\varphi}+be^{-i\varphi}\big)\tan\left(\frac{\chi}{2}\right)\;.
\end{equation}

The eigenvalue problem \eqref{eq:eigenvalueproblem} on $(\mathbb{S}^2,\bs{q}_+)$ is solved by spherical harmonics $Y_l^m$,
\begin{equation}
    \triangle Y_l^m(\chi,\varphi)=-\tfrac{\Lambda}{3}l(l+1)Y_l^m(\chi,\varphi)\;,
\end{equation}
for which we use the normalization
\begin{equation}
    Y_l^m(\chi,\varphi)=\sqrt{\frac{2l+1}{4\pi}\frac{(l-m)!}{(l+m)!}}P_l^m(\cos{\chi})e^{im\varphi}
\end{equation}
with the orthonormality relation
\begin{equation}
    \int_0^{\pi}\! d\chi'\int_0^{2\pi} \!d\varphi'\,\sin\chi' Y_l^m(\chi',\varphi')\bar{Y}_{l'}^{m'}(\chi',\varphi')=\delta_{ll'}\delta_{mm'}\;.
\end{equation}
Here, $P_l^m$ denote associated Legendre polynomials. Due to the completeness of $Y_l^m$, an arbitrary square-integrable function $f(u,\chi,\varphi)$ on $(\mathbb{S}^2,\bs{q}_+)$ (with an additional dependence on $u$) can be uniquely decomposed as
\begin{equation}
    f(u,\chi,\varphi)=\sum_{l=0}^{\infty}\sum_{m=-l}^l f_l^m(u)Y_l^m(\chi,\varphi)\;,
\end{equation}
where the coefficients $f_l^m(u)$ are obtainable by integration
\begin{equation}
    f_l^m(u)= \int_0^{\pi}\! d\chi'\int_0^{2\pi} \!d\varphi'\,\sin\chi' f(u,\chi',\varphi')\bar{Y}_l^m(\chi',\varphi')\;.
\end{equation}
The last two formulas together give rise to a well-known identity for the Dirac delta distribution at ${\chi=\chi_0}$ and ${\varphi=\varphi_0}$,
\begin{equation}\label{eq:deltadelta}
    \delta(\cos\chi-\cos\chi_0)\delta(\varphi-\varphi_0)=\sum_{l=0}^{\infty}\sum_{m=-l}^l Y_l^m(\chi,\varphi)\bar{Y}_l^m(\chi_0,\varphi_0)\;.
\end{equation}
Thus, for an arbitrary square-integrable source ${\hat{J}=\sum_{l}\sum_m \hat{J}_l^m Y_l^m}$, we can express the particular solution using formula \eqref{eq:Hzerosol},
\begin{equation}\label{eq:HpartgenericJ}
    \hat{H}^{\circ}_{\textrm{part}}(u,\chi,\varphi) =\sum_{l=0}^{\infty}\sum_{m=-l}^l \frac{3/\Lambda}{{-} l(l+1)+2}e^{\mathcal{A}({-}\Lambda l(l+1)/3)}\hat{J}_l^m(u) Y_l^m(\chi,\varphi)\;.
\end{equation}

\begin{figure}
    \centering
    \raisebox{-.5\height}{\includegraphics[width=0.55\textwidth]{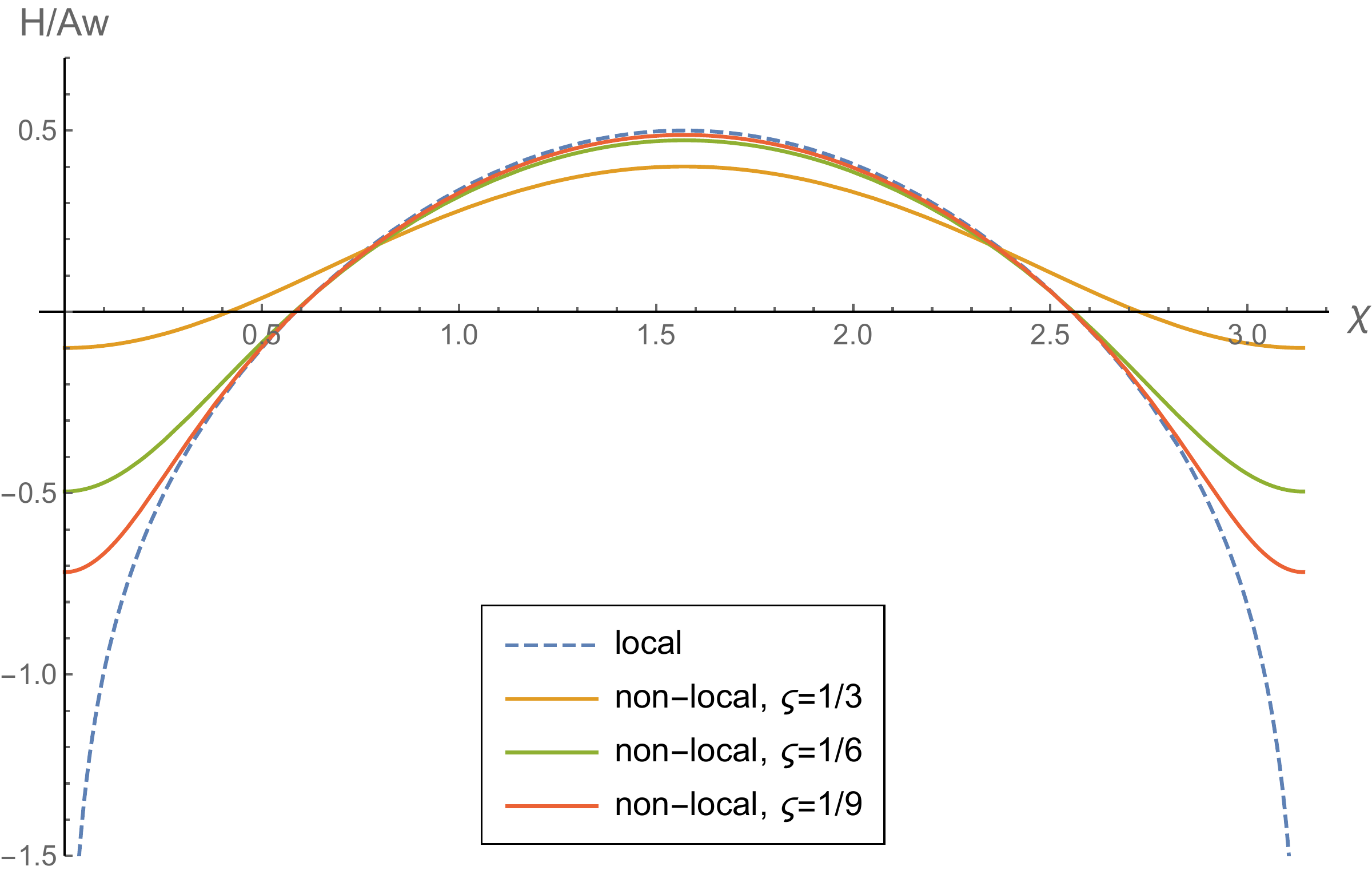}}
    \hspace{.5cm}
    \raisebox{-.5\height}{\includegraphics[width=0.2\textwidth]{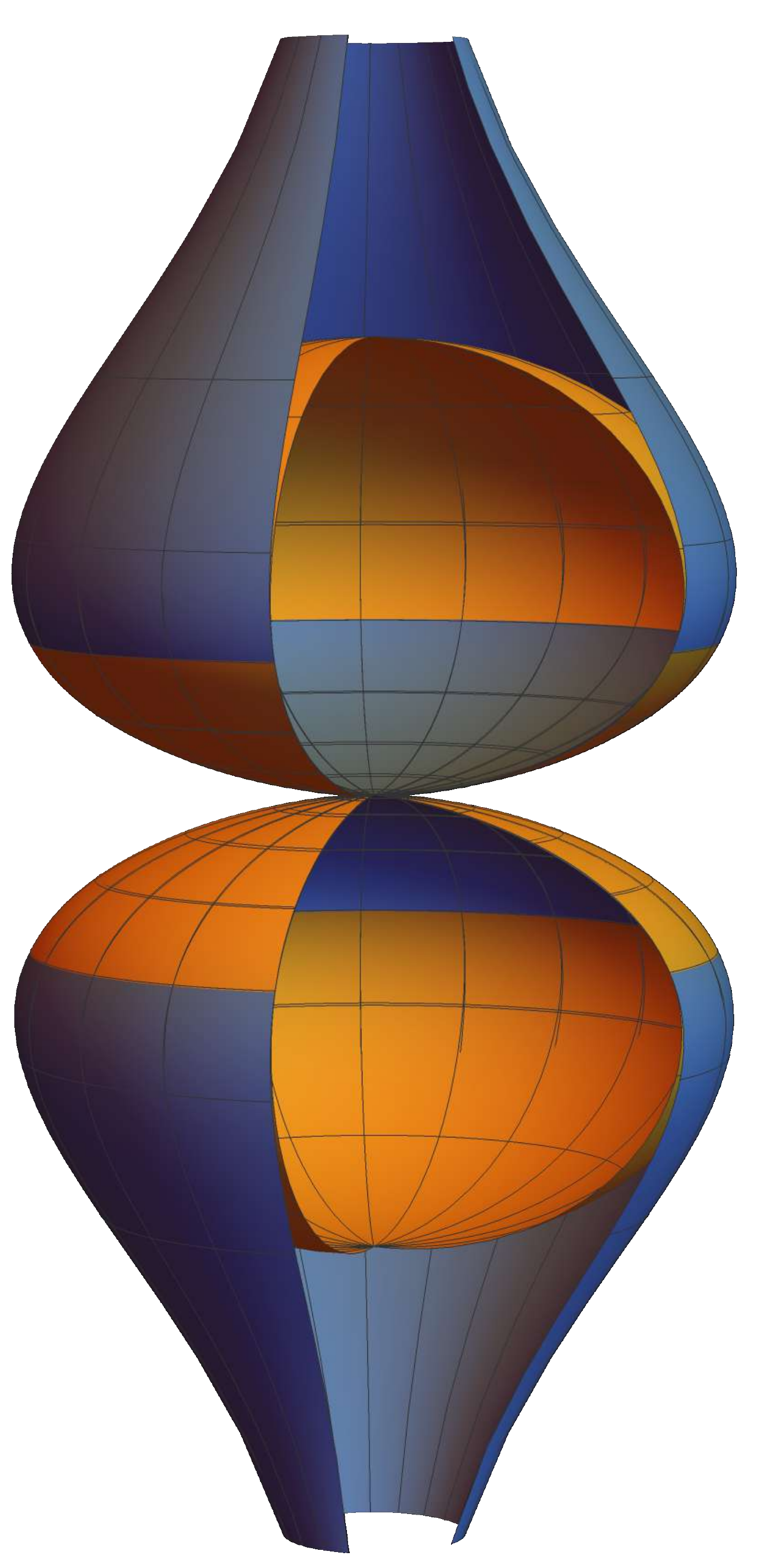}}
    \caption{Non-local analogue of the Hotta--Tanaka solution with ${\Lambda>0}$. Left: A plot of dimensionless functions $\hat{H}^{\circ}_{\textrm{part,loc}}/A w$ (local) and $\hat{H}^{\circ}_{\textrm{part}}/A w$ (non-local) with respect to a dimensionless variable $\chi$. The source is located at both poles ${\chi=0}$ and ${\chi=\pi}$ for three choices of ${\varsigma_{+}\equiv\sqrt{\Lambda/3}\,\ell}$. Right: Three-dimensional representation of the solutions on a sphere with spherical coordinates $\chi$ and $\varphi$. The solutions are characterized through the spherical radius that is given by the values of ${-\hat{H}^{\circ}/Aw+1/2\geq0}$.}
    \label{fig:ht}
\end{figure}

As a simple but important example, let us consider $\hat{J}$ composed of two null sources located at opposite poles of the sphere, ${\chi=0}$ and ${\chi=\pi}$,
\begin{equation}\label{eq:Jtwopart}
    \hat{J}(u,\chi)=\frac{\Lambda A}{6}\left[\delta(\cos\chi-1)+\delta(\cos\chi+1)\right]w(u)\;.
\end{equation}
Here, ${w=w(u)}$ is an arbitrary wave-profile function depending on the null coordinate $u$. In GR, this source induces the \textit{(generalized) Hotta--Tanaka solution with ${\Lambda>0}$} \cite{Hotta:1992qy} (see also \cite{Podolsky:1997ni,Podolsky:1997ri}),\footnote{Authors of \cite{Podolsky:1997ni} use a notation where ${\triangle'\equiv 3\triangle/\Lambda}$, ${G\equiv 2\hat{H}^{\circ}}$, and ${b_0\equiv A}$, in which the field equation takes the form ${(\triangle'+2)G=6\hat{J}/\Lambda}$.
}
\begin{equation}\label{eq:HTpos}
\begin{aligned}
    \hat{H}^{\circ}_{\textrm{part,loc}}(u,\chi)&=-\frac{A}{2}Q_1(\cos\chi)w(u)\;,
    \\
    Q_1(\cos\chi)&=\frac12\cos\chi\log\left(\frac{1+\cos\chi}{1-\cos\chi}\right)-1
    \\
    &=\cos (\chi ) \arctanh (\cos \chi )-1\;,
\end{aligned}
\end{equation}
where $Q_{1}$ denotes the Legendre function of the second kind which is defined as
\begin{equation}
    Q_{1}(x)\equiv\frac{x}{2}\log\left|\frac{1+x}{1-x}\right|-1\;.
\end{equation}
This exact solution of GR has interesting physical and geometric properties: With an impulsive profile ${w=\delta(u)}$ and for ${a=0}$ and ${b=1}$ (appearing in $\tau$ and $Q$, see \eqref{eq:TNKundttauQ}), it can be obtained by taking the \textit{ultra-boost limit} of the de Sitter--Schwarzschild metric where the speed of the source approaches the speed of light and the rest mass approaches zero. Such geometries then represent impulsive spherical gravitational waves in de Sitter background generated by null particles located at poles ${\chi=0}$ and ${\chi=\pi}$ and moving in opposite directions. In order to find an analogous exact solution of the IDG with \eqref{eq:GA} and \eqref{eq:Az}, we have to expand the source \eqref{eq:Jtwopart} as a linear combination of spherical harmonics. Integrating \eqref{eq:deltadelta} over ${\varphi_0}$ and putting ${\chi_0=0}$ and ${\chi_0=\pi}$, we arrive at the relation
\begin{equation}
    \delta(\cos\chi-1)+\delta(\cos\chi+1)=\sum_{k=0}^{\infty} \sqrt{(4k+1)\pi}\, Y_{2k}^0(\chi,\varphi)\;,
\end{equation}
which says that the only non-zero components $\hat{J}_{l}^{m}$ are
\begin{equation}\label{eq:J02k}
    \hat{J}_{2k}^0(u)=\frac{\Lambda A}{3}\sqrt{(4k+1)\pi}\,w(u)\;.
\end{equation}
Upon inserting \eqref{eq:J02k} in \eqref{eq:HpartgenericJ}, we get
\begin{equation}\label{eq:Hhotp}
    \hat{H}^{\circ}_{\textrm{part}}(u,\chi) =-\frac{A}{4}w(u)\sum_{k=0}^{\infty}\frac{4k+1}{(2k+1)k-1}\,e^{-2k(2k+1)\varsigma_{+}^2}P_{2k}(\cos\chi)\;,
\end{equation}
where we introduced a dimensionless parameter ${\varsigma_{+}\equiv\sqrt{\Lambda/3}\,\ell}$. It is unlikely that this infinite series can be resumed to a closed-form expression, but numerical results shows that the sum converges pretty quickly. Thus, we can get an approximate expression with a sufficient precision by truncating the series after first couple of terms. The Hotta--Tanaka solution \eqref{eq:HTpos} can be recovered by taking the limit ${\varsigma_{+}\to 0}$, i.e., ${\ell\ll \sqrt{3/\Lambda}}$ as a direct consequence of the well-known summation identity of Legendre functions of the first kind \cite{Podolsky:1997ni},
\begin{equation}
    \frac12\sum _{k=0}^{\infty } \frac{4 k+1}{(2 k+1)k -1}P_{2 k}(\cos\chi)=Q_{1}(\cos\chi)\;.
\end{equation}
The graphs in Figure~\ref{fig:ht} confirm this limit and show that $\hat{H}^{\circ}_{\textrm{part}}$ is regular at the location of the source. If ${\varsigma_+\to\infty}$, i.e., ${\ell\gg\sqrt{3/\Lambda}}$, then the function $\hat{H}^{\circ}_{\textrm{part}}$ approaches the value ${A w(u)/4}$ everywhere on the sphere. This uniform distribution arises because the source is effectively completely smeared out over $(\mathbb{S}^2,\bs{q}_+)$. The limiting value is non-zero in this case because the area of the sphere is finite. As we will see below this is not true for ${\Lambda\leq0}$, as the 2-spaces of zero $(\mathbb{R}^2,\bs{q}_0)$ and negative $(\mathbb{H}^2,\bs{q}_-)$ constant curvature have an infinite area.

\subsection{Solutions with ${\Lambda<0}$}
If the cosmological constant $\Lambda$ is negative, the 2-space is a hyperbolic plane $\mathbb{H}^2$ equipped with the metric $\bs{q}_-$ of negative Gaussian curvature ${\Lambda/3<0}$. This 2-space can be covered by the \textit{pseudo-spherical coordinates} with the transformation analogous to \eqref{eq:transzetap},\footnote{The Jacobian matrix of \eqref{eq:transzetan} is
\begin{equation*}
    \begin{bmatrix}
    \pp_{\zeta}\chi & \pp_{\bar{\zeta}}\chi
    \\
    \pp_{\zeta}\varphi & \pp_{\bar{\zeta}}\varphi
    \end{bmatrix} = 
    \sqrt{\frac{-\Lambda}{6}}\begin{bmatrix}
    \cosh^2\left(\frac{\chi}{2}\right)e^{-i\varphi} & \cosh^2\left(\frac{\chi}{2}\right)e^{i\varphi}
    \\
    -\frac{i}{2}\coth\left(\frac{\chi}{2}\right)e^{-i\varphi} & \frac{i}{2}\coth\left(\frac{\chi}{2}\right)e^{i\varphi}
    \end{bmatrix}\;.
\end{equation*}
}
\begin{equation}\label{eq:transzetan}
    \zeta=\sqrt{\frac{6}{-\Lambda}} \tanh \left(\frac{\chi}{2}\right) e^{i \varphi}\;,
    \quad
    \chi\in(0,\infty)\;, \quad \varphi\in(0,2\pi)\;,
\end{equation}
in which the metric \eqref{eq:2dmetric} takes form
\begin{equation}
    \bs{q}_-=\frac{3}{-\Lambda}(\bs{\dd} \chi^2+\sinh^2\chi\bs{\dd} \varphi^2)\;.
\end{equation}
Because of evident similarity with \eqref{eq:qplus}, many expressions are just hyperbolic versions of those for positive constant curvature with appropriate sign changes. For instance, the Laplace operator $\triangle$ reads (cf. \eqref{eq:laplpos})
\begin{equation}
    \triangle f =\frac{-\Lambda}{3}\frac{1}{\sinh\chi}\pp_{\chi}\big(\sinh\chi\pp_{\chi}f\big)+\frac{-\Lambda}{3}\frac{1}{\sinh^2\chi}\pp_{\varphi}^2 f\;,
\end{equation}
and the functions ${P}$ and ${Q}$ from \eqref{eq:TNKundtP} and \eqref{eq:TNKundttauQ} are given by (cf. \eqref{eq:PQpos})
\begin{equation}
    P=\frac{1}{\cosh^2\left(\frac{\chi}{2}\right)}\;,
    \quad
    Q=a\frac{\cosh{\chi}}{\cosh^2\left(\frac{\chi}{2}\right)}+\sqrt{\frac{6}{-\Lambda}}\big(\bar{b}e^{i\varphi}+be^{-i\varphi}\big)\tanh\left(\frac{\chi}{2}\right)\;.
\end{equation}

The eigenvalue problem of the Laplace operator on the hyperbolic plane is significantly more difficult than on the sphere. Thus, it is very convenient to use the characterization of the particular solution of the IDG with \eqref{eq:GA} and \eqref{eq:Az} by means of the heat kernel method \eqref{eq:solheatker} instead of eigenfunction expansion \eqref{eq:Hzerosol}. Fortunately, an explicit expression for the heat kernel on $(\mathbb{H}^2,\bs{q}_-)$ is well known \cite{mckean1970},\footnote{See also \cite{chavel1984eigenvalues} for more details on the heat equation and eigenfunctions of the Laplace operator on the hyperbolic plane.}
\begin{equation}\label{eq:heatkerH}
    K(s;\mathrm{x},\mathrm{x}')=\sqrt{\frac{-\Lambda}{3}}\frac{\sqrt{2}}{(4\pi s)^{\frac32}}e^{s\Lambda/12}\int_{\varrho}^{\infty}\!\!d\tilde{\varrho}\,\frac{\tilde{\varrho}\,e^{-\tilde{\varrho}^2/4s}}{\Big[\cosh\Big(\sqrt{\frac{-\Lambda}{3}}\,\tilde{\varrho}\Big)-\cosh\Big(\sqrt{\frac{-\Lambda}{3}}\,\varrho\Big)\Big]^{\frac12}}\;.
\end{equation}
This formula depends only on the geodesic distance ${\varrho\equiv|\mathrm{x}-\mathrm{x}'|}$ between two points ${\mathrm{x},\mathrm{x'}\in \mathbb{H}^2}$, which can be expressed in our coordinate system as
\begin{equation}\label{eq:geoddist}
    \varrho=\sqrt{\frac{3}{-\Lambda}}\arccosh\big(\cosh\chi\cosh\chi'-\cos(\varphi-\varphi')\sinh\chi\sinh\chi'\big)
    \;.
\end{equation}
The particular solution induced by an arbitrary source $\hat{J}$ is obtained from \eqref{eq:solheatker},
\begin{equation}\label{eq:HsolneggeneralbJ}
    \hat{H}^{\circ}_{\textrm{part}}(u,\chi,\phi)=-e^{-2\ell^2\Lambda/3}\int_{\ell^2}^{\infty} \!ds \,e^{2s\Lambda/3}\!\!\!\int_0^{\infty}\! d\chi'\int_0^{2\pi} \!d\varphi' \frac{3}{-\Lambda}\sinh\chi' K(s;\chi,\varphi,\chi',\varphi')\hat{J}(u,\chi')\;,
\end{equation}
whenever the integral converges.

\begin{figure}
    \centering
    \raisebox{-.5\height}{\includegraphics[width=0.55\textwidth]{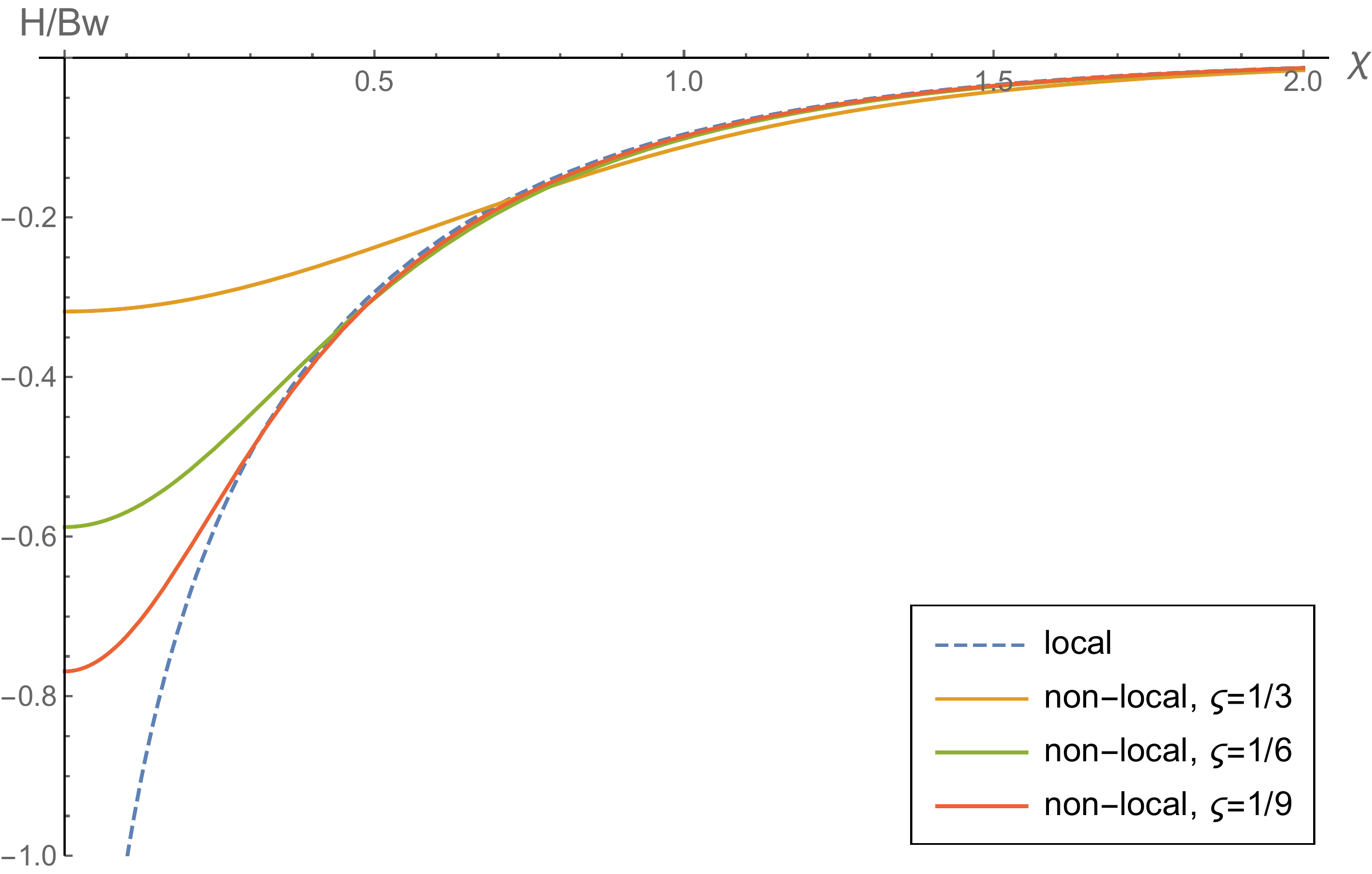}}
    \hspace{.5cm}
    \raisebox{-.5\height}{\includegraphics[width=0.4\textwidth]{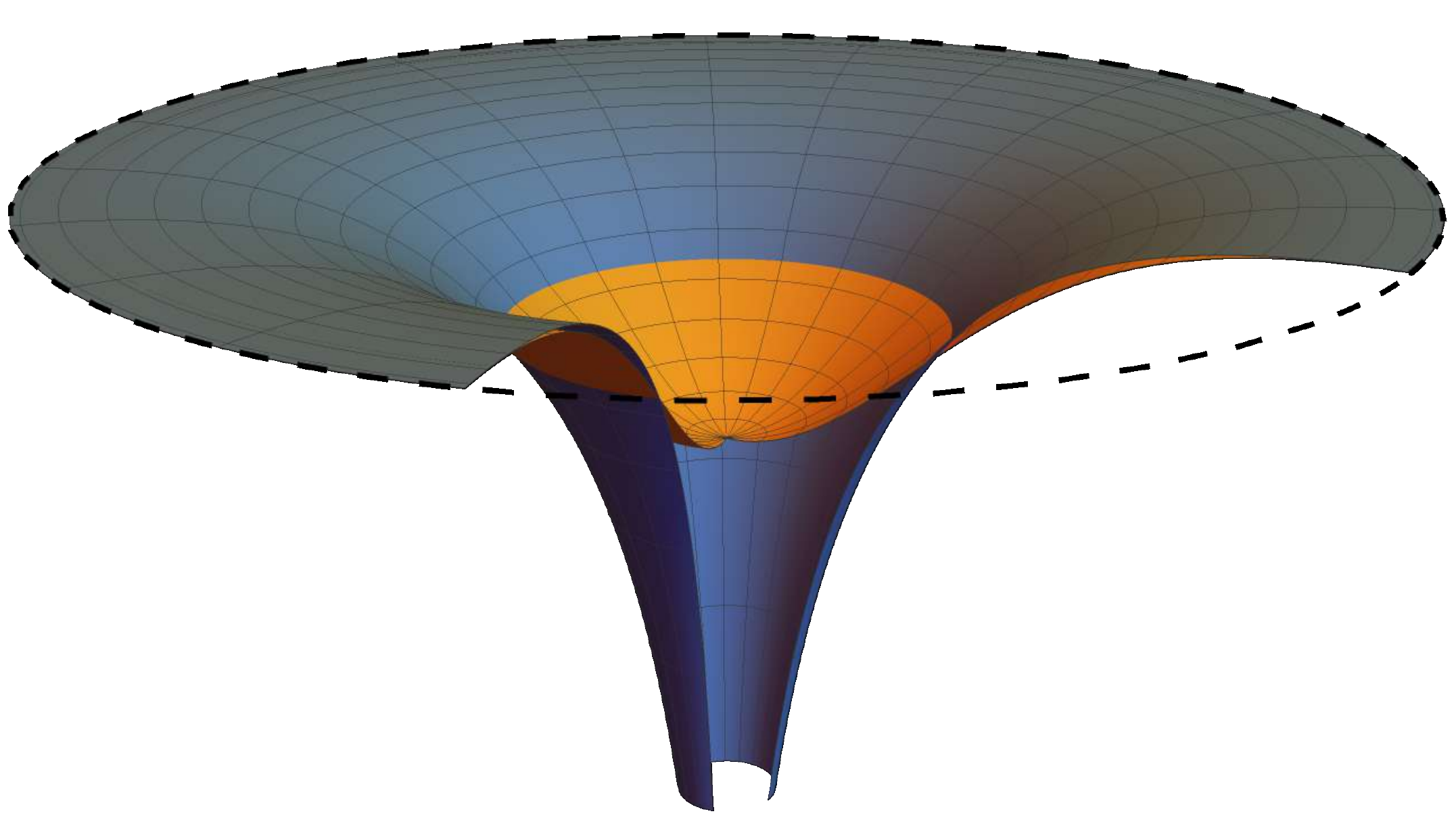}}
    \caption{Non-local analogue of the Hotta--Tanaka solution with ${\Lambda<0}$. Left: A plot of dimensionless functions $\hat{H}^{\circ}_{\textrm{part,loc}}/B w$ (local) and $\hat{H}^{\circ}_{\textrm{part}}/B w$ (non-local) with respect to a dimensionless variable $\chi$ for three choices of ${\varsigma_{-}\equiv\sqrt{-\Lambda/3}\,\ell}$. The source is located at the origin ${\chi=0}$. Right: Three-dimensional representation of the solutions on the Poincar\'e disc model of a hyperbolic plane in pseudo-spherical coordinates $\chi$ and $\varphi$, which is given by compactifying function $\tanh(\chi/2)$. The values of $\hat{H}^{\circ}/Bw$ are drawn on the vertical axis (perpendicular to the disc). The dashed circle marks the conformal infinity, ${\chi=\infty}$.}
    \label{fig:htn}
\end{figure}

Let us now consider a source located at origin of the pseudo-spherical coordinates of the hyperbolic plane, ${\chi=0}$,
\begin{equation}\label{eq:oneparticleneg}
    \hat{J}(u,\chi)=\frac{-\Lambda B}{6}\delta(\cosh\chi-1)w(u)\;,
\end{equation}
which, in GR, generates the \textit{(generalized) Hotta--Tanaka solution with ${\Lambda<0}$} \cite{Hotta:1992qy} (see also \cite{Podolsky:1997ni,Podolsky:1997ri}),
\begin{equation}\label{eq:HottaTanakaneg}
\begin{aligned}
    \hat{H}^{\circ}_{\textrm{part,loc}}(u,\chi)&=-\frac{B}{2}Q_1(\cosh\chi)w(u)\;,
    \\
    Q_1(\cosh\chi)&=\frac12\cosh \chi \log \left(\frac{\cosh\chi+1}{\cosh\chi-1}\right)-1
    \\
    &=-\cosh \chi \log \left(\tanh \left(\frac{\chi}{2}\right)\right)-1\;.
\end{aligned}
\end{equation}
Complementary to the previous case, this geometry with the impulsive profile ${w=\delta(u)}$ and for ${a=0}$ and ${b=1}$, is a result of the ultra-boost limit of the anti-de Sitter--Schwarzschild metric. It represents hyperboloidal impulsive gravitational waves generated by a null particle located at the origin ${\chi=0}$ and propagating in anti-de Sitter spacetime. To find a corresponding non-local solution, we insert the source \eqref{eq:oneparticleneg} in the formula \eqref{eq:HsolneggeneralbJ} and realize that for ${\chi'=0}$ the geodesic distance $\varrho$ is independent of $\varphi'$. Hence, we can integrate over both coordinates ${\chi'}$ and ${\varphi'}$ due to the relation ${\delta(\chi')=\delta(\cosh\chi'-1)\sinh\chi'}$. After interchanging the order of integration, we obtain
\begin{equation}
    \hat{H}^{\circ}_{\textrm{part}}(u,\chi)
    =-\sqrt{\frac{-\Lambda}{6}}\frac{B}{4\sqrt{\pi}}e^{-2\ell^2\Lambda/3}w(u)\int_{\sqrt{\frac{3}{-\Lambda}}\,\chi}^{\infty}\!\!d\tilde{\varrho}\,\frac{\tilde{\varrho}}{\Big[\cosh\Big(\sqrt{\frac{-\Lambda}{3}}\,\tilde{\varrho}\Big)-\cosh\chi\Big]^{\frac12}}\int_{\ell^2}^{\infty} \!ds \,\frac{e^{3s\Lambda/4-\tilde{\varrho}^2/4s}}{s^{\frac32}}\;,
\end{equation}
where we used the expressions for \eqref{eq:heatkerH} and \eqref{eq:geoddist} evaluated at ${\chi'=0}$. The inner integration can be performed analytically,
\begin{equation}
    \int_{\ell^2}^{\infty} \!ds \,\frac{e^{3s\Lambda/4-\tilde{\varrho}^2/4s}}{s^{\frac32}}=\frac{\sqrt{\pi }}{\tilde{\varrho}}  \left[e^{-\frac{1}{2} \sqrt{-3\Lambda } \tilde{\varrho}} \left(\erf\left(\frac{\tilde{\varrho}- \sqrt{-3\Lambda } \ell ^2}{2 \ell }\right)+1\right)+e^{\frac{1}{2} \sqrt{-3\Lambda } \tilde{\varrho}} \left(\erf\left(\frac{ \sqrt{-3\Lambda } \ell ^2+\tilde{\varrho}}{2 \ell }\right)-1\right)\right]\;,
\end{equation}
but the remaining integral probably does not have a closed-form expression. By changing the variable of integration ${\tilde{\chi}\equiv\sqrt{-\Lambda/3 }\,\tilde{\varrho}}$ and denoting ${\varsigma_{-}\equiv\sqrt{-\Lambda/3}\,\ell}$, we can write the solution in the compact form\footnote{A similar source in the half-space Poincar\'e-type coordinates was studied in \cite{Dengiz:2020xbu} by a direct eigenfunction expansion. Due to a quite complicated and very different form of this solution, it is rather difficult to compare it to a much simpler solution \eqref{eq:Hhottaneg} for the source \eqref{eq:oneparticleneg} in the pseudo-spherical coordinates (and arbitrary ${a}$ and $b$).}
\begin{equation}\label{eq:Hhottaneg}
    \hat{H}^{\circ}_{\textrm{part}}(u,\chi) =-\frac{B}{4 \sqrt{2}}\, e^{2 \varsigma_{-} ^2}w(u)\! \int_{\chi}^{\infty}\!\!d\tilde{\chi}\,\frac{e^{\frac{3 \tilde{\chi}}{2}} \left[\erf\left(\frac{\tilde{\chi}+3 \varsigma_{-} ^2}{2 \varsigma_{-} }\right)-1\right]+e^{-\frac{3 \tilde{\chi}}{2}} \left[\erf\left(\frac{\tilde{\chi}-3 \varsigma_{-} ^2}{2 \varsigma_{-} }\right)+1\right]}{\sqrt{\cosh \tilde{\chi}-\cosh \chi}}\;,
\end{equation}
which can be evaluated numerically. In the local limit, ${\varsigma_{-}\to 0}$, i.e., ${\ell\ll \sqrt{-3/\Lambda}}$, or far from the source ${\chi\gg\ell}$, we recover the Hotta--Tanaka solution \eqref{eq:HottaTanakaneg} due to the integral identity,
\begin{equation}
   \frac{1}{\sqrt{2}} \int_{\chi}^{\infty}\!\!d\tilde{\chi}\,\frac{e^{-\frac{3 \tilde{\chi}}{2}}}{\sqrt{\cosh \tilde{\chi}-\cosh \chi}}= Q_1(\cosh\chi)\;,
\end{equation}
The regularity of $\hat{H}^{\circ}_{\textrm{part}}$ at the location of the source as well as the described limiting behavior are also evident from the graphs in Figure~\ref{fig:htn}. The function $\hat{H}^{\circ}_{\textrm{part}}$ vanishes in the limit ${\varsigma_-\to\infty}$, i.e., ${\ell\gg\sqrt{-3/\Lambda}}$, since the source is completely smeared over the hyperbolic plane, which has an infinite area.

\subsection{Solutions with ${\Lambda=0}$}

Finally, we discuss the case of vanishing cosmological constant ${\Lambda}$, where the 2-space is a standard Euclidean plane $\mathbb{R}^2$ with the metric ${\bs{q}_0}$ of zero Gaussian curvature. The coordinate transformation to the \textit{polar coordinates}\footnote{The Jacobian matrix of \eqref{eq:transzeta0} is
\begin{equation*}
    \begin{bmatrix}
    \pp_{\zeta}\chi & \pp_{\bar{\zeta}}\chi
    \\
    \pp_{\zeta}\varphi & \pp_{\bar{\zeta}}\varphi
    \end{bmatrix} = 
    \frac{1}{\sqrt{2}}\begin{bmatrix}
    e^{-i\varphi} & e^{i\varphi}
    \\
    -\frac{i}{\chi}e^{-i\varphi} & \frac{i}{\chi}e^{i\varphi}
    \end{bmatrix}\;.
\end{equation*}
}
\begin{equation}\label{eq:transzeta0}
    \zeta=\frac{\chi}{\sqrt{2}}e^{i \varphi}\;,
    \quad
    \chi\in(0,\infty)\;, \quad \phi\in(0,2\pi)
\end{equation}
transforms the flat metric \eqref{eq:2dmetric} to
\begin{equation}\label{eq:q0}
    \bs{q}_0=\bs{\dd} \chi^2+\chi^2\bs{\dd} \varphi^2\;.
\end{equation}
The Laplace operator $\triangle$ takes the form
\begin{equation}
    \triangle f =\frac{1}{\chi}\pp_{\chi}\left(\chi\pp_{\chi}f\right)+\frac{1}{\chi^2}\pp_{\varphi}^2 f
\end{equation}
and the functions $P$ and $Q$ from \eqref{eq:TNKundtP} and \eqref{eq:TNKundttauQ} are
\begin{equation}
    P=1\;,
    \quad
    Q=a+\frac{\chi}{\sqrt{2}}\big(\bar{b}e^{i\varphi}+b e^{-i\varphi}\big)\;.
\end{equation}

In what follows we assume that ${\tau\neq 0}$, however, the derivation is exactly the same for the special case ${\tau= 0}$, which is obtained by replacing ${\hat{J}\to J}$ and ${\hat{H}^{\circ}_{\textrm{part}}\to H^{\circ}_{\textrm{part}}}$. As before, we use the heat kernel method \eqref{eq:solheatker} to find a particular solution of the IDG with \eqref{eq:GA} and \eqref{eq:Az}. The heat kernel on the Euclidean plane is given by the well-known formula
\begin{equation}
    K(s;\mathrm{x},\mathrm{x}')=\frac{e^{-\varrho^2/4s}}{4\pi s}\;,
\end{equation}
where ${\varrho\equiv|\mathrm{x}-\mathrm{x}'|^2}$ denotes the distance between two points ${\mathrm{x},\mathrm{x}'\in \mathbb{R}^2}$, which in polar coordinates reads
\begin{equation}
    \varrho =\chi^2+\chi'^2-2\chi\chi'\cos(\varphi-\varphi')\;.
\end{equation}
Using \eqref{eq:solheatker}, we can express the particular solution for an arbitrary source $\hat{J}$ in the form
\begin{equation}\label{eq:arbsourcesol}
    \hat{H}^{\circ}_{\textrm{part}}(u,\chi,\varphi)=-\int_{\ell^2}^{\infty} \!\! ds\!  \int_0^{\infty}\! d\chi'\int_0^{2\pi} \!d\varphi'\,\chi'K(s;\chi,\varphi,\chi',\varphi')\hat{J}(u,\chi')\;.
\end{equation}

As before, we take a null source located at the origin of the polar coordinates of the Euclidean plane, ${\chi=0}$, which is described by
\begin{equation}\label{eq:singlepsource}
    \hat{J}(u,\chi)=\frac{C}{2\pi\chi}\delta(\chi)w(u)\;.
\end{equation}
In GR, such a source produces the \textit{(generalized) Aichelburg--Sexl solution} \cite{Aichelburg:1970dh},
\begin{equation}\label{eq:aichsexl}
    \hat{H}^{\circ}_{\textrm{part,loc}}(u,\chi)=\frac{C}{2 \pi } \log \left(\chi/\chi_1 \right)w(u)\;.
\end{equation}
The corresponding metric with an impulsive profile ${w=\delta(u)}$ for ${\tau=0}$ (${a=1}$ and ${b=0}$) is obtained by boosting the Schwarzschild metric to the speed of light with vanishing rest mass. It corresponds to a plane impulsive gravitational wave generated by a null particle located at the origin ${\chi=0}$. In order to find a non-local counterpart to this solution, we should insert the source \eqref{eq:singlepsource} in \eqref{eq:arbsourcesol}. Unfortunately, this method leads to a diverging integral (even in the local case). Following \cite{Frolov:2015usa,Boos:2018bxf}, one can avoid this problem by deducing the solution $\hat{H}^{\circ}_{\textrm{part}}$ from the higher-dimensional results with the help of recursive relation
\begin{equation}\label{eq:recursform}
    \hat{H}^{\circ}_{\textrm{part},n}(u,\chi)=-2\pi\int_{\chi_1}^{\chi}\!d\tilde{\chi}\, \hat{H}^{\circ}_{\textrm{part},n+2}(u,\tilde{\chi})\tilde{\chi}\;.
\end{equation}

\begin{figure}
    \centering
    \raisebox{-.5\height}{\includegraphics[width=0.55\textwidth]{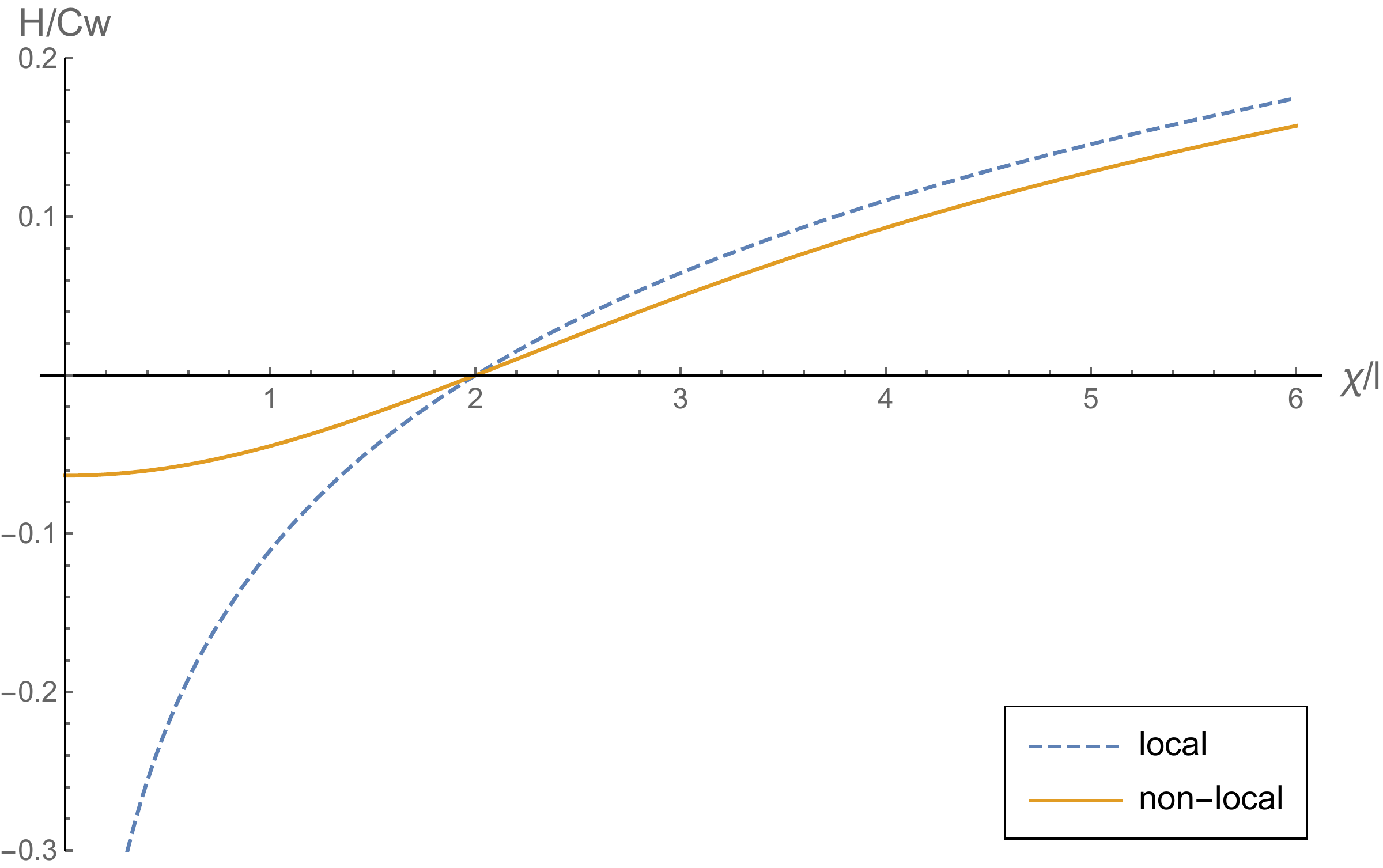}}
    \hspace{.5cm}
    \raisebox{-.5\height}{\includegraphics[width=0.4\textwidth]{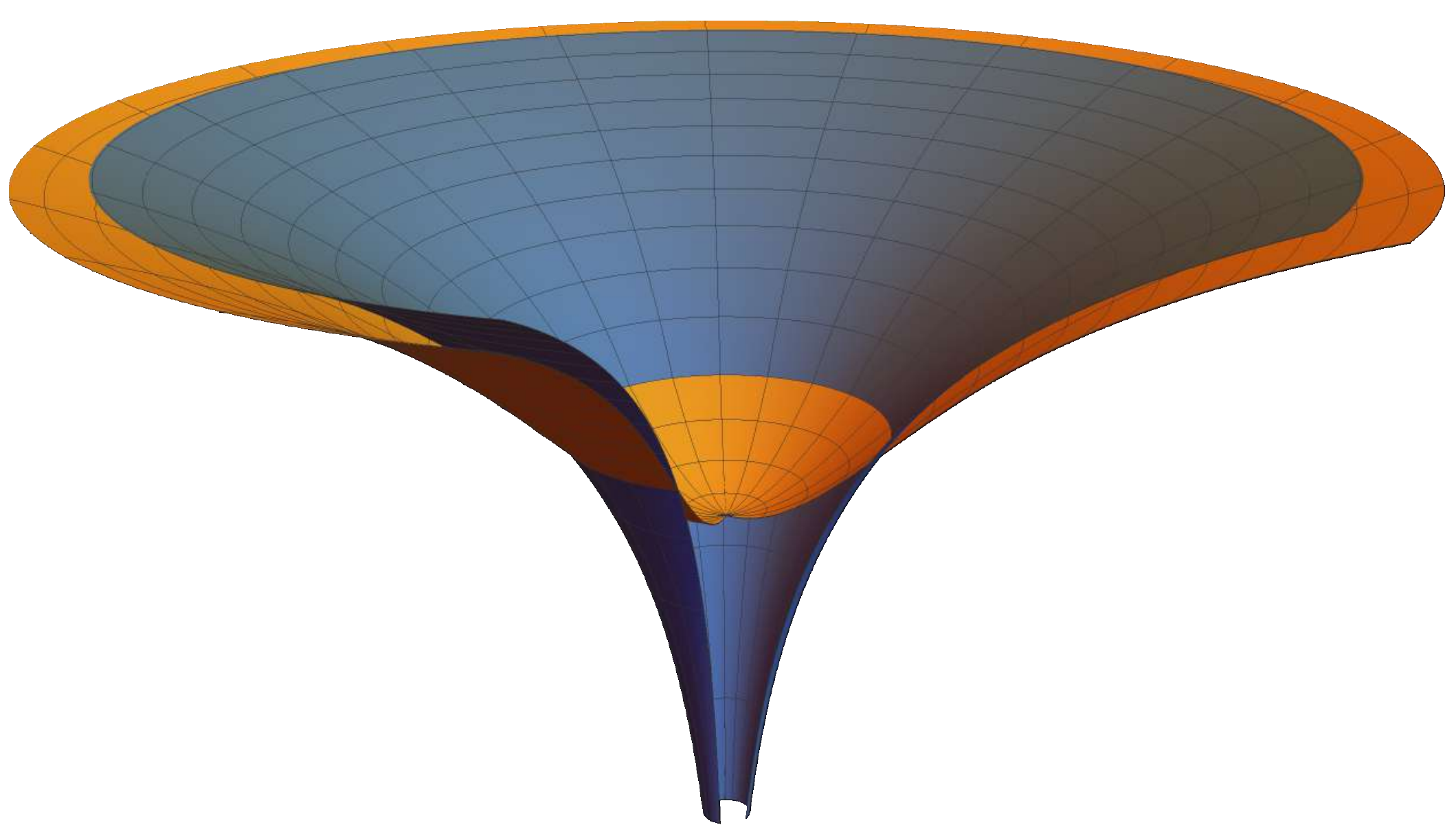}}
    \caption{Non-local analogue of the Aichelburg--Sexl solution. Left: A plot of dimensionless functions $\hat{H}^{\circ}_{\textrm{part,loc}}/C w$ (local) and $\hat{H}^{\circ}_{\textrm{part}}/C w$ (non-local) with respect to a dimensionless variable ${\chi/\ell}$ for the choice ${\chi_1/\ell=2}$. The source is located at the origin ${\chi=0}$. Right: Three-dimensional representation of the solutions on a Euclidean plane in polar coordinates $\chi$ and $\varphi$, where the values of $\hat{H}^{\circ}/Cw$ are drawn on the vertical axis.}
    \label{fig:as}
\end{figure}

By a straightforward generalization of \eqref{eq:arbsourcesol} to the $n$-dimensional Euclidean space, where the heat kernel is
\begin{equation}
    K(s;\mathrm{x},\mathrm{x}')=\frac{e^{-\varrho^2/4s}}{(4\pi s)^{n/2}}\;,
\end{equation}
one can find that the integrals converge for all ${n>2}$ and we get
\begin{equation}
    \hat{H}^{\circ}_{\textrm{part},n}(u,\chi)=-C w(u)\int_{\ell^2}^{\infty} \!\! ds \,\frac{e^{-\chi^2/4s}}{(4\pi s)^{n/2}}=-\frac{1}{4} C \pi ^{-\frac{n}{2}} \chi^{2-n} \left(\Gamma \left(\frac{n}{2}-1\right)-\Gamma \left(\frac{n}{2}-1,\frac{\chi^2}{4 \ell ^2}\right)\right)\;.
\end{equation}
Thus, the solution for ${n=2}$ can be extrapolated using the recursive formula \eqref{eq:recursform},\footnote{The special case, ${\tau=0}$ (${a=1}$ and ${b=0}$), is a known solution, which was originally found in the linearized theory \cite{Frolov:2015usa}, where authors studied a null-particle collision, and later rediscovered as an exact solution in \cite{Kilicarslan:2019njc}.}
\begin{equation}\label{eq:Haich}
    \hat{H}^{\circ}_{\textrm{part}}(u,\chi)=\frac{C}{4 \pi }\left[2\log \left(\chi/\chi_1\right) -\Ei \left(-\chi^2/4\ell^2\right) +\Ei \left(-\chi_1^2/4\ell^2\right) \right]w(u)\;.
\end{equation}
This expression reduces back to the Aichelburg--Sexl solution \eqref{eq:aichsexl} in the local limit ${\ell\to 0}$ or far from the source ${\ell\ll\chi}$ (up to a constant shift given by $\chi_1$, which is of no physical relevance \cite{Boos:2018bxf}). This limiting behavior and the regularity of $\hat{H}^{\circ}_{\textrm{part}}$ at the location of the source can be seen from the graphs in Figure~\ref{fig:as}. Similar to ${\Lambda<0}$, the function $\hat{H}^{\circ}_{\textrm{part}}$ approaches zero in the limit ${\ell\to\infty}$ because the Euclidean plane has an infinite area.

\section{Curvature}\label{sc:CU}
It turns out that the functions $\hat{H}^\circ_\text{part}$ of the IDG solutions \eqref{eq:Hhotp}, \eqref{eq:Hhottaneg}, and \eqref{eq:Haich} are non-singular at the locations of the sources. A natural question that arises is whether these solutions are regular or not. To answer this question, one has to investigate the spacetime curvature. Since both GR as well as IDG solutions belong to the class of almost universal spacetimes, all polynomial scalar curvature invariants are necessarily constant. This means that they are free of the \textit{scalar curvature singularities}. However, such spacetimes could in principle still possess the so-called \textit{non-scalar curvature singularities} (see, e.g., \cite{Geroch:1968ut,Ellis:1977pj}), as it seems to be the case of the GR solutions at the locations of the sources. To confirm the absence of the non-scalar curvature singularities in the IDG solutions, one should investigate all possible timelike and null geodesics passing through the locations of the sources and check that the curvature tensor components in parallel-propagated frames along such geodesics remain bounded. This ensures that all physically relevant observers experience finite tidal forces and there are no obstructions in the extension of geodesics past the potentially problematic points (e.g, locations of the sources, coordinate singularities). Even if we restrict ourselves to finding a single instance of such geodesics, such a task is rather involved and goes beyond the scope of this paper. Instead, we will show that the non-locality regularizes the curvature components expressed in the natural null frame \eqref{eq:naturalframe}, which indicates that the non-scalar curvature singularity could be ameliorated.

Since $\hat{H}^\circ_\text{part}$ is axially symmetric and proportional to the profile function $w$ for all solutions, we introduce an auxiliary function $h$ such that 
\begin{equation}
    \hat{H}^\circ_\text{part}(u,\chi)=w(u)h(\chi)\;.
\end{equation}
For simplicity we will also take ${W^{\circ}=0}$, which follows automatically from \eqref{eq:V0} for ${\Lambda\neq0}$. In the case ${\Lambda > 0}$, the transverse 2-space is a sphere $\mathbb{S}^2$ equipped with the metric $\bs{q_+}$ given by \eqref{eq:qplus}. The null source is located at the opposite poles of the sphere, ${\chi = 0}$ and ${\chi = \pi}$, as prescribed by $\hat{J}$ of the form \eqref{eq:Jtwopart}. The only non-vanishing components of the Ricci and Weyl tensors in the natural null frame read
\begin{equation}
    \Phi_{22} = \frac{\Lambda}6 \frac{Q}{P} w(u) (h'' + \cot\chi h' + 2h)\;,
    \quad
    \Psi_4 = \frac{\Lambda}6 \frac{Q}{P} w(u) (h'' - \cot\chi h')\;,
    \quad
    0<\chi<\pi\;.
\end{equation}
Substituting the local Hotta--Tanaka solution $\hat{H}^\circ_\text{part,loc}$ from \eqref{eq:HTpos}, we get\footnote{Note that the expressions are valid only outside the sources, ${0<\chi<\pi}$. In fact, $\Phi_{22}^{\textrm{loc}}$ contains Dirac delta terms at ${\chi=0}$ and ${\chi=\pi}$ as follows directly from the field equations \eqref{eq:Gphi22J} (with ${\mathcal{G}(\mathcal{D})=1}$).}
\begin{equation}
    \Phi_{22}^\text{loc} = 0\;,
    \quad
    \Psi_4^\text{loc} = - A \frac{\Lambda}6 \frac{Q}{P} w(u) \sin^{-2}\chi\;,
\end{equation}
where we can see that the component of the Weyl tensor diverges at the poles ${\chi = 0}$ and ${\chi = \pi}$.
The components of the curvature for the non-local solution $\hat{H}^\circ_\text{part}$ from \eqref{eq:Hhotp} cannot be expressed in a closed-form. However, their limits at the locations of the sources,
\begin{equation}\label{eq:PhiPsilimP}
    \begin{aligned}
        &\lim_{\chi \rightarrow 0} \Phi_{22}^{\text{nonloc}} = - \lim_{\chi \rightarrow \pi} \Phi_{22}^{\text{nonloc}} = A \frac{\Lambda}{12} a w(u) \sum_{k=0}^\infty (4k+1) e^{-2k(2k+1) \varsigma_{+}^2}\;,\\
        &\lim_{\chi \rightarrow 0} \Psi_4^{\text{nonloc}} =  \lim_{\chi \rightarrow \pi} \Psi_4^{\text{nonloc}} = 0
    \end{aligned}
\end{equation}
are finite because the series in the first line is convergent.

Consider now the case $\Lambda = 0$, where the transverse 2-space is the Euclidean plane $\mathbb{R}^2$ with the flat metric ${\bs{q}_0}$ given by \eqref{eq:q0}. The source $\hat{J}$ of the form \eqref{eq:singlepsource} is located at the origin ${\chi = 0}$ in the polar coordinates. The non-vanishing curvature tensor components are given by
\begin{equation}
    \Phi_{22} = \frac12 Q w(u) (h'' + \chi^{-1} h')\;, 
    \quad 
    \Psi_4 = \frac12 Q w(u) (h'' - \chi^{-1} h')\;,
    \quad
    \chi>0\;.
\end{equation}
Plugging in the local Aichelburg--Sexl solution $\hat{H}^\circ_\text{part,loc}$ from \eqref{eq:aichsexl}, the component of the Weyl tensor obviously blows up at the source position,
\begin{equation}
    \Phi_{22}^{\text{loc}} = 0, \quad \Psi_4^\text{loc} = -\frac{C}{2\pi\chi^2} Q w(u)\;.
\end{equation}
The Ricci and Weyl tensor components can be expressed explicitly also for the non-local solution $\hat{H}^\circ_\text{part}$ given by \eqref{eq:Haich},
\begin{equation}
    \Phi_{22}^{\text{nonloc}} = C Q w(u) \frac{e^{-\frac{\chi^2}{4\ell^2}}}{8\pi \ell^2}\;, 
    \quad
    \Psi_4^\text{nonloc} = \frac{C}{8\pi\ell^2\chi^2} Q w(u) e^{-\frac{\chi^2}{4\ell^2}} (\chi^2 + 4(1 - e^{\frac{\chi^2}{4\ell^2}})\ell^2)\;.
\end{equation}
Moreover, it turns out that their limits at the origin ${\chi = 0}$ are finite,
\begin{equation}\label{eq:PhiPsilim0}
    \lim_{\chi \rightarrow 0} \Phi_{22}^{\text{nonloc}} = \frac{C a w(u)}{8\pi\ell^2}\;,
    \quad
    \lim_{\chi \rightarrow 0} \Psi_4^\text{nonloc} = 0\;.
\end{equation}

Finally, let us discuss the case ${\Lambda < 0}$ for which the transverse 2-space is the hyperbolic plane $\mathbb{H}^2$ equipped with the metric $\bs{q}_-$ from \eqref{eq:oneparticleneg}. The source $\hat{J}$ is now given by \eqref{eq:oneparticleneg} and it is located at the origin ${\chi = 0}$ in the pseudo-spherical coordinates. The components of the Ricci and Weyl tensor in terms of the function $\hat{H}^\circ$ are analogous to the case of positive $\Lambda$ with the trigonometric functions being replaced by the hyperbolic ones together with some additional sign changes,
\begin{equation}
    \Phi_{22} = -\frac{\Lambda}6 \frac{Q}{P} w(u) (h'' + \coth\chi h' - 2h)\;, 
    \quad 
    \Psi_4 = -\frac{\Lambda}6 \frac{Q}{P} w(u) (h'' - \coth\chi h')\;,
    \quad
    \chi>0\;.
\end{equation}
For the local Hotta--Tanaka solution $\hat{H}^\circ_\text{part,loc}$ from \eqref{eq:HottaTanakaneg}, the components reduce to
\begin{equation}
    \Phi_{22}^\text{loc} = 0, \quad \Psi_4^\text{loc} = B \frac{\Lambda}6 \frac{Q}{P} w(u) \sinh^{-2}\chi\;.
\end{equation}
Unfortunately, for the non-local solution with $\hat{H}^\circ_\text{part}$ given by \eqref{eq:Hhottaneg}, we were not able to express the components of the Ricci and Weyl tensor in a closed form, neither to evaluate the limits at the location of the source analytically. Nevertheless, the numerical integration suggests that limits are finite also in this case,
\begin{equation}\label{eq:PhiPsilimN}
    \lim_{\chi \rightarrow 0} \Phi_{22}^{\text{nonloc}} = -B \frac{\Lambda}{3} a w(u) \times \text{const}\;,
    \quad
    \lim_{\chi \rightarrow 0} \Psi_4^{\text{nonloc}} = 0\;.
\end{equation}

Notice that in all three cases the components of the Weyl tensor approaches zero at the locations of the sources for the IDG solutions. A similar property was pointed out for the spherical solutions of the linearized IDG \cite{Buoninfante:2018rlq}; here we extend this observation to the exact IDG solutions with null sources. Obviously, our conclusion should be considered only as a hint, because the natural null frame might possibly be ill-behaved. As mentioned above, further investigation using parallel-propagated frames along timelike and null geodesics is needed.

Remark that in the limit of large non-local scale, ${\ell\to\infty}$, one can immediately see that $\Psi_4^{\text{nonloc}}$ vanishes everywhere (since $h$ is zero for ${\Lambda\le0}$ and constant for ${\Lambda>0}$), so the spacetimes become conformally flat. Furthermore, ${\Phi_{22} = A \Lambda Q w(u)/12P}$ for ${\Lambda>0}$ in this regime. This is also reminiscent to the observation of \cite{Buoninfante:2018rlq} noticing that the Bertotti--Robinson spacetime (conformally flat spherically symmetric metric that is a direct product of the 2-dimensional anti de-Sitter and the 2-dimensional sphere) is a solution of the highly non-local regime of IDG (a theory obtained by neglecting the Einstein--Hilbert term in the action), which can be understood as the limit ${\ell\to\infty}$.


\section{Conclusions}\label{sc:C}
In this work, we have studied the role of the almost universal spacetimes in the non-local theory of gravity with an infinite number of derivatives. Thanks to some remarkable properties of these spacetimes, it turns out that the field equations of IDG can be reduced considerably. In particular, we have obtained a single non-local but linear differential equation \eqref{eq:redfielddeltatilde}, which contains only the Laplace operator $\triangle$ on 2-dimensional spaces of constant curvature. In this sense, the class of almost universal spacetimes represents an ideal metric ansatz from the mathematical as well as physical point of view. The reason is because the \textit{linearity} of the resulting field equation makes the non-local problem much more tractable and its \textit{non-locality} makes the solutions strongly dependent on $\mathcal{F}_i(z)$ at all values of $z$. Indeed, we have shown that such a non-local linear equation can be solved exactly either by eigenfunction decomposition of the source or using the heat kernel method (for the natural exponential form-factor $\exp(-\ell^2\triangle)$). 

Both of these mathematical techniques are easily applicable to any source because the eigenfunctions of $\triangle$ and the heat kernels are well known on the spaces of constant curvature. To demonstrate these methods on specific examples, we have derived the non-local versions of the (generalized) Hotta--Tanaka and Aichelburg--Sexl solutions \eqref{eq:Hhotp}, \eqref{eq:Hhottaneg}, and \eqref{eq:Haich}, which describe gravitational waves generated by null sources propagating in de Sitter, anti-de Sitter, and Minkowski spacetime, respectively. Far away from the sources or for small values of the length scale of non-locality $\ell$, these spacetimes reduce to the GR solutions.

We have also briefly discussed the regularity of the obtained solutions. Figures \ref{fig:ht}, \ref{fig:htn}, and \ref{fig:as}, indicates that the regularization occurs due to the non-locality for the functions $\hat{H}^{\circ}_{\textrm{part}}$. Belonging to the class of almost universal spacetimes, both GR and IDG solutions are free from scalar curvature singularities, since all scalar curvature invariants are constant. We have shown that the components of the curvature tensors expressed in the natural null frames, which blow up for the GR solutions, remain finite in IDG, see \eqref{eq:PhiPsilimP}, \eqref{eq:PhiPsilim0}, and \eqref{eq:PhiPsilimN}. This hints at the absence of non-scalar curvature singularities, although a proper verification of this hypothesis, which we leave for a future work, requires a rather non-trivial construction of frames parallel-propagated along all timelike and null geodesics. Furthermore, we found that the Weyl tensor components approach zero at the location of the source and vanish everywhere in the limit ${\ell\to\infty}$ (the spacetime becomes conformally flat). Both properties resemble similar results for the spherically symmetric solutions in the linearized IDG or for the highly non-local regime of IDG (where the Einstein--Hilbert term is neglected) \cite{Buoninfante:2018rlq}.

An interesting continuation of our work would be to consider spinning null sources that are usually referred to as \textit{gyratons} \cite{Frolov:2005in}. Counterparts of these exact GR solutions \cite{Bonnor:1970sb,Griffiths1972} (see also \cite{Griffiths:2009dfa}), have been recently discovered in the linearized IDG \cite{Boos:2020ccj}. Thus, it would be natural to examine whether they can be promoted to exact solutions of the full IDG. Although the gyratonic spacetimes still belong to the Kundt class, the function ${W^{\circ}}$ is no longer independent of $\zeta$. As a consequence, this line of research requires generalization of the class of almost universal spacetimes beyond rank-2 tensors $\bs{B}$ whose trace is of type N, because the spinning sources typically introduce a non-vanishing component $\Phi_{12}$ of the Ricci tensor.


\section*{Acknowledgements}

We would like to thank to Vojt\v{e}ch Pravda (Prague) for inspiration to study exact solution of IDG in the context of almost universal spacetimes and Ji\v{r}\'i Podolsk\'y (Prague) for explaining details of the properties of Kundt geometries. I.K. and A.M. were supported by Netherlands Organization for Scientific Research (NWO) grant no. 680-91-119. T.M. acknowledges the support of the Czech Science Foundation GA\v{C}R grant no. GA19-09659S.


\appendix


\section{Identities for TNS spacetimes}\label{app:SCid}

To derive the identities \eqref{eq:SCid}, we frequently employ the commutator of the covariant derivatives
\begin{equation}
    [\nabla_b,\nabla_a] T_{c_1\cdots c_k} = T_{d\cdots c_k} R^d{}_{c_1 ab} + \dots + T_{c_1 \cdots d} R^d{}_{c_k ab}
\end{equation}
and the fact that mixed terms and terms quadratic in $\bs{\nabla}^k\bs{S}$ and $\bs{\nabla}^k\bs{C}$, ${k\geq 0}$ do not contribute to rank-2 tensors for TNS spacetimes.
The first identity follows from the application of the commutator to the outermost covariant derivatives
\begin{equation}
    \nabla_c \nabla_b \square^n S^{ca} = \frac{R}{3} \square^n S^a{}_b + \nabla_b \nabla_c \square^n S^{ca}\;,
\end{equation}
where $\nabla_c \square^{n} S^{ca}$ vanishes since $\bs{\nabla}^k\bs{S}$, $k\geq 0$ do not contribute to rank-1 tensors.

The second identity can be obtained by moving $\square$ in front of the two outermost covariant derivatives ,
\begin{equation}
    \nabla_c\nabla_d\square^n C_b{}^{cda} = \left( \square + \frac{R}{3} \right) \nabla_c\nabla_d\square^{n-1} C_b{}^{cda} - \frac{R}{3} \nabla_c\nabla_d\square^{n-1} C_b{}^{acb} = \left( \square + \frac{R}{3} \right)^{\!n} \nabla_c\nabla_d C_b{}^{cda}\;,
\end{equation}
where
\begin{equation}
    \nabla_c\nabla_d\square^n C_a{}^{bcd} = 0\;,
\end{equation}
as can be shown using the cyclic symmetry of the Weyl tensor and by sorting the covariant derivatives. Finally, substituting
\begin{equation}
    \nabla_d\nabla_c C_a{}^{cbd} = \frac{1}{2} \left( \square - \frac{R}{3} \right) S_a{}^b\;,
\end{equation}
which follows from the divergence of the contracted Bianchi identity, we arrive at the second identity.


\section{Newman--Penrose formalism}\label{app:NP}

Following the Newman--Penrose (NP) formalism, we introduce an orthonormal null frame consisting of two real null vectors $\bs{l}$, $\bs{n}$, a complex null vector $\bs{m}$ and its conjugate $\bar{\bs{m}}$ satisfying\footnote{In the index-free notation, we use $\cdot$ to denote contractions of adjacent indices. For example, the contraction of a 1-form $\bs{\alpha}$ with a vector $\bs{v}$ reads ${\bs{\alpha}\cdot\bs{v}=\alpha_a v^a}$. We also employ the musical isomorphism to raise indices of a 1-forms ${\bs{\alpha}^{\sharp}=\bs{g}^{-1}\cdot\bs{\alpha}}$, and lower indices of vectors ${\bs{v}^{\flat}=\bs{g}\cdot\bs{v}}$.}
\begin{equation}
    \bs{l} \cdot \bs{n}^\flat = -1, \quad \bs{m} \cdot \bar{\bs{m}}^\flat = 1\;.
\end{equation}
The covariant derivative can be then expressed in terms of the directional derivative operators $\mathrm{D}$, $\Delta$, $\delta$ and $\bar\delta$ as
\begin{equation}
    \bs{\nabla} = - \bs{n}^\flat \mathrm{D} - \bs{l}^\flat \Delta + \bar{\bs{m}}^\flat \delta + \bs{m}^\flat \bar\delta\;.
\end{equation}
The directional derivatives of the frame vectors can be written using 12 complex NP spin coefficients, which are denoted by lower-case Greek letters,
\begin{equation}
\begin{aligned}
    \mathrm{D}\bs{l} &= (\varepsilon + \bar\varepsilon) \bs{l} - \bar\kappa \bs{m} - \kappa \bar{\bs{m}}\;,
    &
    \mathrm{D} \bs{n} &= -(\varepsilon + \bar\varepsilon) \bs{n} + \pi \bs{m} + \bar\pi \bar{\bs{m}}\;,
    &
    \mathrm{D} \bs{m} &= \bar\pi \bs{l} - \kappa \bs{n} + (\varepsilon - \bar\varepsilon) \bs{m}\;,
    \\
    \Delta \bs{l} &= (\gamma + \bar\gamma) \bs{l} - \bar\tau \bs{m} - \tau \bar{\bs{m}}\;,
    &
    \Delta \bs{n} &= -(\gamma + \bar\gamma) \bs{n} + \nu \bs{m} + \bar\nu \bar{\bs{m}}\;,
    &
    \Delta \bs{m} &= \bar\nu \bs{l} - \tau \bs{n} + (\gamma - \bar\gamma) \bs{m}\;,
    \\
    \delta \bs{l} &= (\bar\alpha + \beta) \bs{l} - \bar\rho \bs{m} - \sigma \bar{\bs{m}}\;,
    &
    \delta \bs{n} &= -(\bar\alpha + \beta) \bs{n} + \mu \bs{m} + \bar\lambda \bar{\bs{m}}\;,
    \\
    \delta \bs{m} &= \bar\lambda \bs{l} - \sigma \bs{n} - (\bar\alpha - \beta) \bs{m}\;,
    &
    \bar\delta \bs{m} &= \bar\mu \bs{l} - \rho \bs{n} + (\alpha - \bar\beta) \bs{m}\;.
\end{aligned}
\end{equation}


\section{Parallel-propagated frames in Kundt spacetimes}\label{app:ppframe}

In the Newman--Penrose (NP) formalism (see Appendix \ref{app:NP}), the geometric definition of the Kundt spacetimes  \eqref{eq:Kundtconds} translates to
\begin{equation}\label{eq:KundtcondsNP}
    \varepsilon + \bar\varepsilon = \kappa = 0\;, \quad \sigma = 0\;, \quad \rho = - (\theta + i \omega) = 0\;,
\end{equation}
where the shear scalar corresponds directly to the spin coefficient $\sigma$.
Moreover, it is always possible to transform the frame to achieve
\begin{equation}\label{eq:taucond}
    \tau = \bar\alpha + \beta\;,
\end{equation}
while retaining \eqref{eq:KundtcondsNP}. As a consequence, the covariant derivative of the Kundt null vector $\bs{l}$ reduces to the form
\begin{equation}
    \nabla_a l_b = - (\gamma + \bar\gamma) l_a l_b + 2 \bar\tau l_{(a} m_{b)} + 2 \tau l_{(a} \bar{m}_{b)}\;.
\end{equation}

For Kundt metrics \eqref{eq:Kundt}, one can introduce a natural null frame
\begin{equation}\label{eq:naturalframe}
    \bs{l} = \bs{\pp}_r, \quad \bs{n} = \bs{\pp}_u - H \bs{\pp}_r, \quad \bs{m} = P \bs{\pp}_{\bar\zeta} -  P\bar{W}\bs{\pp}_r\;,
\end{equation}
and its dual coframe
\begin{equation}
    \bs{l}^{\flat} = - \bs{\dd}u, \quad \bs{n}^{\flat} = -\bs{\dd}r - H \bs{\dd}u - W \bs{\dd}\zeta - \bar{W} \bs{\dd}\bar\zeta, \quad \bs{m}^{\flat} = P^{-1} \bs{\dd}\zeta\;.
\end{equation}
Obviously, the null vector $\bs{l}$ satisfies \eqref{eq:KundtcondsNP}, i.e.\ it is geodesic and affinely parametrized with vanishing expansion, shear, and twist. Moreover, ${\varepsilon = \bar\varepsilon = 0}$, so that the directional derivative $\mathrm{D}$ of the frame vectors reduce to
\begin{equation}
    \mathrm{D} \bs{l} = 0, \quad \mathrm{D} \bs{n} = \pi \bs{m} + \bar\pi \bar{\bs{m}}, \quad \mathrm{D} \bs{\bs{m}}= \bar\pi \bs{l}\;,
\end{equation}
where
\begin{equation}
    \pi = - \frac{1}{2} P W_{,r}\;.
\end{equation}
One can check that the condition \eqref{eq:taucond} is met and
\begin{equation}
    \tau = \frac{1}{2} P \bar{W}_{,r}\;.
\end{equation}

The frame \eqref{eq:naturalframe} is not parallel-propagated along $\bs{l}$. However, we can employ null rotations with $\bs{l}$ fixed
\begin{equation}
    \bs{l}' = \bs{l}, \quad \bs{m}' = \bs{m} + N \bs{l}, \quad \bs{n}' = \bs{n} + N \bs{\bar{m}} + \bar{N} \bs{m} + N\bar{N} \bs{l},
\end{equation}
where $N$ is a complex parameter, to set $\pi$ to zero. The spin coefficients $\kappa$, $\varepsilon$, $\rho$, $\sigma$, and $\tau$ are invariant with respect to this transformation and the condition \eqref{eq:taucond} remains satisfied. The spin coefficient $\pi$
transforms as
\begin{equation}
    \pi' = \pi + \mathrm{D}\bar{N}\;,
\end{equation}
therefore we require ${\bar{N} = \frac{1}{2} P W + \bar{N}_0(u,\zeta,\bar\zeta)}$ where $\bar{N}_0$ is an arbitrary function independent of $r$. Choosing ${N_0 = 0}$, one obtains the parallel-propagated frame
\begin{equation}\label{eq:ppframe}
    \bs{l} = \bs{\pp}_r, \quad \bs{n} = \bs{\pp}_u - \left(H + \frac{3}{4} P^2 W\bar{W}\right) \bs{\pp}_r + \frac{1}{2} P^2 \bar{W} \bs{\pp}_\zeta + \frac{1}{2} P^2 W \bs{\pp}_{\bar\zeta}, \quad \bs{m} = P \bs{\pp}_{\bar\zeta} - \frac{1}{2} P\bar{W}\bs{\pp}_r\;,
\end{equation}
whose dual coframe is
\begin{equation}
    \bs{l}^{\flat} = - \bs{\dd}u, \quad \bs{n}^{\flat} = -\bs{\dd}r - \left(H + \frac{1}{4}P^2 W\bar{W}\right) \bs{\dd}u - \frac{1}{2} W \bs{\dd}\zeta - \frac{1}{2} \bar{W} \bs{\dd}\bar\zeta, \quad \bs{m}^{\flat} = P^{-1} \bs{\dd}\zeta - \frac{1}{2} P \bar{W} \bs{\dd}u\;.
\end{equation}
The change of frame only affects the component $\Psi_4$ of the Weyl tensor,
\begin{equation}
    \Psi'_3 = \Psi_3, \quad \Psi'_4 = \Psi_4 + 4 \bar{N} \Psi_3\;.
\end{equation}


\section{Weyl type III TNS Kundt spacetimes with ${\tau \neq 0}$}\label{app:tnskundttaunz}

Kundt spacetimes of Weyl type III and traceless Ricci type N are TN spacetimes, i.e., all symmetric rank-2 tensors $\bs{B}$ constructed from the metric, the Riemann tensor, and its covariant derivatives of an arbitrary order take the form \eqref{eq:TN}. Moreover, if the condition \eqref{eq:F2_NP}, ${C^\text{II}_{ab}=0}$, is satisfied, then such spacetimes are TNS and the traceless part of the rank-2 tensors $\bs{B}$ reduces to \eqref{eq:TNS}. Notice that \eqref{eq:F2_NP} is met for ${\tau = 0}$ or if the Weyl type specializes to type N, ${\Psi_3 = 0}$. For genuine Weyl type III with ${\tau \neq 0}$, the condition ${C^\text{II}_{ab}=0}$ can be rewritten as
\begin{equation}\label{eq:C2cond}
    \left[Q\tau(Q\tau W^\circ)_{,\zeta}\right]^2 + \left[Q\bar\tau(Q\bar\tau \bar{W}^\circ)_{,\bar\zeta}\right]^2 = 0\;,
\end{equation}
where we substituted for $\Psi_3$ from  \eqref{eq:phi22}. The term in the first square bracket is independent of $\bar\zeta$, whereas the term in the second square bracket is independent of $\zeta$. Therefore, necessarily 
\begin{equation}
\label{eq:TNSWcircDE}
    Q\tau(Q\tau W^\circ)_{,\zeta} = c\;, 
\end{equation}
where $c$ is a complex function of $u$ subject to
\begin{equation}
    c^2 + \bar{c}^2 = 0\;,
\end{equation}
which implies ${\Re c= \pm\Im c}$ (or equivalently, ${c=|c|\, e^{i(\frac{\pi}{4}+\frac{\pi}{2}k)}}$, ${k\in\mathbb{Z}}$).

The solution of \eqref{eq:TNSWcircDE} has to be discussed according to the value of $\Lambda$ and the expression
\begin{equation}
    \varkappa \equiv \frac{\Lambda}{6} a^2 + b \bar{b}\;.
\end{equation}
The sign of $\varkappa$ is invariant under coordinate transformations preserving the form of the Kundt metric \eqref{eq:Kundt}, and thus it can be used to distinguish subfamilies of the Kundt class of metrics \cite{Ozsvath:1985qn} (see also \cite{Griffiths:2009dfa}). If ${\Lambda \neq 0}$ and ${\varkappa \neq 0}$, \eqref{eq:TNSWcircDE} has a general solution
\begin{equation}
    W^\circ=\frac{1}{Q\tau} \left[\sqrt{\frac{6}{\Lambda}} \frac{c}{\sqrt{\varkappa}} \arctanh\left(\sqrt{\frac{\Lambda}{6}} \frac{a+\bar{b} \zeta}{\sqrt{\varkappa}}\right) + c_1\right]\;,
\end{equation}
where $c_1$ is an arbitrary function of $u$. Note that the special case when both $\Lambda$ and $\kappa$ vanish is excluded by the assumption $\tau \neq 0$. For $\Lambda = 0$, \eqref{eq:TNSWcircDE} implies
\begin{equation}
    W^\circ=\frac{c}{b^2}\zeta+c_1\;.
\end{equation}
Finally, for $\varkappa = 0$, we obtain
\begin{equation}
    W^\circ=\frac{1}{Q\tau} \left( c \frac{a-\bar{b}\zeta}{\bar{b}(b + \frac{\Lambda}{6} \bar{b}\zeta^2)} + c_1 \right)\;.
\end{equation}

Plugging \eqref{eq:TNSWcircDE} back to \eqref{eq:phi22}, it follows for TNS Kundt spacetimes with $\tau \neq 0$ that $\Psi_3$ is given by
\begin{equation}
    \Psi_3 = - \frac{P^2 c}{Q^2 \tau}\;.
\end{equation}
It can be also convenient to rewrite $V$ from \eqref{eq:Ntaunonzero} appearing in $\Phi_{22}$ as
\begin{equation}
    V = 2 P^2 \left(2\tau\bar\tau + \frac{\Lambda}{3}\right) W^\circ\bar{W}^\circ + 2 \frac{P^3}{Q^2} \left( \frac{c \bar{W}^\circ}{\tau} + \frac{\bar{c} W^\circ}{\bar\tau} \right)\;,
\end{equation}
where we employed \eqref{eq:TNSWcircDE} in the form
\begin{equation}
    W^\circ_{,\zeta} = \frac{c}{(Q\tau)^2} - (\log Q\tau)_{,\zeta} W^\circ
\end{equation}
and the identity
\begin{equation}
    2 P_{,\zeta} - P(\log Q\tau)_{,\zeta} = -\frac{\Lambda}{3\tau}
\end{equation}
whose validity can be proved straightforwardly by substituting $P$, $Q$, and $\tau$.


\bibliography{references.bib}

\end{document}